\documentclass[journal]{IEEEtran}

\IEEEoverridecommandlockouts                              % This command is only needed if 
\usepackage{amsmath} % assumes amsmath package installed

\usepackage{amssymb}
\usepackage{latexsym}
\usepackage{multirow}
\usepackage{cite}
\usepackage{mathtools}
 \usepackage{relsize}
\usepackage{soul}
\usepackage{color}
\usepackage{physics}
\usepackage{comment}
\usepackage{algorithm2e}

\newcommand*{\qed}{\hfill\ensuremath{\square}}

\newtheorem{Assum}{Assumption}
\newtheorem{Lem}{Lemma}
\newtheorem{Rem}{Remark}
\newtheorem{Cor}{Corollary}

\begin{document}
\title{Rapid  Transitions  with  Robust Accelerated Delayed Self Reinforcement for Consensus-based
Networks }
\author{Anuj Tiwari\thanks{Preprint submitted to IEEE Transactions on Control Systems Technology (Oct. 2020). A. Tiwari is with the Mechanical Engineering Department, U. of Washington, Seattle,
WA 08195-2600 USA e-mail: anujt@uw.edu.}~and~Santosh~Devasia, \IEEEmembership{Fellow, IEEE}\thanks{S. Devasia, Fellow,~IEEE, is with the Mechanical Engineering Department, U. of Washington, Seattle,
WA 08195-2600 USA e-mail: devasia@uw.edu (see http://faculty.washington.edu/devasia/).}
}

\maketitle

\begin{abstract}
Rapid transitions  are important for quick response of consensus-based, multi-agent networks to external stimuli. 
While high-gain can increase response speed,  potential instability  tends to limit the maximum possible gain, and therefore, 
limits the maximum convergence rate to consensus during transitions.   Since the update law for multi-agent networks with 
symmetric graphs can be considered 
as the gradient of its Laplacian-potential function, Nesterov-type accelerated-gradient approaches from  optimization theory, can further improve the
convergence rate of such networks. An advantage of the 
accelerated-gradient approach is that it can be implemented  using  accelerated delayed-self-reinforcement  (A-DSR),  which  does  not  require  new information  from  the network  nor modifications  in  the network connectivity.
However, the accelerated-gradient approach is not directly applicable to 
general directed  graphs since the update law is not the 
gradient of the Laplacian-potential function. 
The main contribution of this work is to extend the accelerated-gradient approach to general directed graph networks, without requiring the graph to be strongly connected. 
Additionally, while both the momentum term and outdated-feedback term in the
accelerated-gradient approach are important in general, it is shown that the momentum term alone is sufficient to achieve balanced robustness and rapid transitions without oscillations in the dominant mode, for networks whose graph Laplacians have real spectrum. 
Simulation results are presented to illustrate the performance improvement  with the proposed Robust A-DSR of $40\%$ in structural robustness and  $50\%$ in convergence rate to consensus, 
when compared to the case without the A-DSR. Moreover, experimental results  are presented that show a similar 
$37\%$ faster convergence  with the
 Robust A-DSR when compared to the case without  the  A-DSR. 
\end{abstract}


\section{INTRODUCTION}
\label{intro_section}

The performance of consensus-based, 
multi-agent networks, such as the response to external stimuli, depends on 
rapidly transitioning from one operating point (consensus value) to another, e.g.,  in  flocking of natural systems,~\cite{Huth_92,Vicsek_95}, as well as engineered systems such as autonomous vehicles, swarms of robots, e.g.,~\cite{Jadbabaie_03,Ren_Beard_05,Olfati_Saber_06} 
and other  networked systems such as aerospace control~\cite{Mark_aero_2019} microgrids~\cite{Cucuzzella_19_Microgrids,Schiffer_16_microgrids}, flexible structures~\cite{Naiming_19_flex_structures}.
Rapid cohesive transitions, e.g., in the orientation of the agents from one consensus value to another, is seen in flocking behaviour during predator attacks and migration ~\cite{Ioannou_12, nagy2010hierarchical}. 
Thus, there is interest to increase the convergence rate
to consensus for such networked multi-agent systems.

\vspace{0.1in}
There  is a fundamental limit to the achievable rate of convergence  using existing neighbor-based update laws
for a given network, e.g., of the form 
\begin{align}
\hat{X}[k+1] &   
=
\hat{X}[k] + u[k]
=  \hat{X}[k] -  \alpha  L \hat{X}[k],
\label{system_eq}
\end{align}
where the 
current state is $\hat{X}[k]$, the updated state is $\hat{X}[k+1]$,  $\alpha$ is the update gain, 
$L$ is the graph Laplacian, and $k$ represents the time instants $t_k = k \delta_t$ with $\delta_t$ as the sampling time-period. 
The 
convergence rate depends on the  eigenvalues of the matrix $P = \left(\textbf{I}- \alpha L\right)$ \cite{fagnani2017introduction}, which in turn depends on the 
eigenvalues of the graph Laplacian $L$. 
For example, if the underlying graph  is undirected and connected, it is well known that convergence to consensus can be achieved provided the update gain $\alpha$ is  
sufficiently small, e.g.,~\cite{Olfati_Murray_07}. 
The update gain can be selected to maximize the convergence rate, and typically, 
a larger gain $\alpha$ tends to increase the convergence rate. 
Nevertheless, for a given graph  (i.e., a given graph Laplacian $L$), 
the range of the acceptable update gain $\alpha$ 
is limited, which in turn, limits the achievable rate of convergence~\cite{Devasia_2019_IJC}.
Typically, the convergence rate tends to be slow if the number of agent inter-connections is small compared to the number of agents, e.g.,~\cite{Carli_08}. 
{ Faster convergence can be achieved using 
randomized time-varying connections,} as shown in,   e.g.,~\cite{Carli_08}. The update sequence  of the agents can also be arranged to improve convergence, e.g.,~\cite{Fanti_15}.
The problem is that the graph connectivity might be fixed and therefore the  Laplacian $L$ cannot be varied over time. In such cases, with a fixed Laplacian $L$, 
the need to maintain stability limits the range of acceptable update gain $\alpha$, and therefore, 
limits the rate of convergence. 
This convergence-rate limitation motivates ongoing efforts to develop  new approaches to improve the network performance, e.g.,~\cite{fast_convergence_16_duan}. Furthermore, in addition to convergence-rate, an important consideration is robustness of the approach,  e.g., as studied in~\cite{Zhenhong_robust_19, Montijano_robust_15}.

\vspace{0.1in}
Since the  neighbor-based update ($u$ in Eq.~\eqref{system_eq}) can be obtained from the gradient of the 
Laplacian potential 
$\Phi_{{\cal{G}}} = \hat{X}^T L \hat{X} $ for undirected graphs, i.e., 
$ u  = -(\alpha / 2)\nabla \Phi_{{\cal{G}}}$, 
Nestertov-type accelerated approaches, used to speed up  gradient-based optimization~\cite{Rumelhart_86,QIAN1999145,Nesterov_83,Jakovetic_Moura_2014,VanScoy_Lynch_18},  can be used to  improve the 
convergence rate.
Previous works have considered the
use of some parts of the  accelerated gradients (from  optimization theory)  for graph-based multi-agent networks. 
For example, the addition of a momentum term 
(of the form $ \hat{X}[k] -  \hat{X}[k-1],  $ as in, e.g., ~\cite{Rumelhart_86}) in the update law  has been shown to improve  the response speed under update-bandwidth limits~\cite{Devasia_2018_JDSMC,Devasia_2019_IJC}. These works have also shown that the use of such reinforcement can lead to  non-diffusive,  wave-like response propagation seen in natural systems such as bird flocks~\cite{Attanasi_14}.
Similarly, the addition of a  Nesterov term without the momentum term, also referred to as an outdated-feedback  (of the form
 $ L(\hat{X}[k] -  \hat{X}[k-1]) $,   as in e.g.,~\cite{Nesterov_83}), has been shown to result in  faster convergence in~\cite{Cao_ren_2010,Moradian_19}, and to enable a linear rate of convergence using a time-varying gain  in~\cite{Bu_2018}.   Time-varying gains, however,  require a global resetting
 of each agent's gain at start of each transition, which might not be always feasible because the 
 start of a transition might not be known to all agents. 
  { 
The combination of  both, the momentum term and the outdated-feedback term,  can further improve the  convergence rate of consensus-based networks when compared to the use of either term alone~\cite{Devasia_ICPS_2019,Devasia_20,tiwari2019cohesive}. 
}
Note that an advantage of  such  accelerated-gradient-based approach is that the update can be implemented by 
using an accelerated delayed-self-reinforcement  (A-DSR), where each agent only uses 
current and past information from the network. This use of already existing information is advantageous since the 
convergence improvement is achieved without the need to change the network connectivity 
and without the need for additional information from the network.
Nevertheless, the update law for more general graphs with non-symmetric Laplacian (e.g., general directed graphs) 
cannot be obtained from the gradient of the 
graph potential~\cite{OlfatiSaberIEEETAC_04,Zhang_hui_IJC_2015}.  
Gradients along local agent-wise potential have been considered along with weight-balancing to improve the performance for directed graphs, 
e.g.,~\cite{Makhdoumi_2015,Khan_Xin_2020}.
However, these approaches rely on the graph being strongly connected, which excludes applications such as platoons where overall information flow between two agents is not bi-directional over the  graph. 
Therefore, the current  Nesterov-based approach and its stability analysis  cannot be directly applied for general  directed graphs (which are not strongly connected),  which are addressed in the current work.

\vspace{0.1in}
The main contribution of this article is to  design a 
Nesterov-type  accelerated update for  general  graph networks 
using a local potential function for each agent.
However, since the resulting update law does not necessarily reduce the overall Laplacian potential~\cite{Zhang_hui_IJC_2015}, the convergence studies from optimization methods cannot be used to establish stability~\cite{Jakovetic_Moura_2014,VanScoy_Lynch_18}. 
Moreover, while Lyapunov functions can be found to study stability for general directed graphs~\cite{Zhang_hui_IJC_2015},  the gradient of these Lyapunov functions does not lead to the control update law, and hence accelerated methods cannot be directly applied using these Lyapunov functions.  
Prior methods that use either the momentum term alone or the outdated-feedback term alone also do not address the stability when both terms are used for general directed  graphs. In this context, a  contribution of this article is to  develop stability conditions for the proposed  generalized  accelerated  approach, with both the momentum and outdated-feedback terms.
The current article 
expands on our  prior work in ~\cite{Devasia_ICPS_2019}, which used a fixed  ratio between the momentum and outdated-feedback terms,
{by (i)~proposing  the general  case with varying ratios between the momentum and outdated-feedback terms; (ii)~developing a stability condition for the generalized approach, (iii)~designing the A-DSR to achieve fast response while maximizing structural robustness, (iv) illustrating the importance of momentum term over the outdated-feedback term for graph networks with real spectrum and (v)~presenting experimental results to comparatively evaluate  the performance, with and without A-DSR.}

\vspace{0.1in}
The article begins by presenting the structurally-robust,  convergence-rate improvement problem, along with the limits of  standard consensus-based   update in Section~\ref{problem_formulation}. 
The proposed A-DSR based approach is introduced in Section~\ref{Section_General_ADSR_approach}, and the stability conditions of the A-DSR approach are developed in Section~\ref{Section_stability_proofs}, followed by the derivation of analytical Robust A-DSR approach for maximizing robustness in Section~\ref{subsection_robust_convergence_with_ADSR}. 
 Section~\ref{simulation-section}  comparatively evaluates the performance with and without A-DSR through simulations, and Section~\ref{experimental-section} presents experimental results. Lastly, conclusions from the article are reported in Section~\ref{conclusion-section}.

\vspace{0.1in}
\section{Problem formulation}
\label{problem_formulation}
\vspace{-0.01in}
This section introduces graph-based consensus dynamics used to model networked systems, and describes the convergence limits with structural robustness achievable due to stability bounds on the update gain in standard neighbor-based consensus dynamics. Finally, the problem statement of the article is stated.

\vspace{0.1in}
\subsection{Background: graph-based control}
\label{Section_graph_based_control}

Let the multi-agent network be modeled using a graph representation, where 
the connectivity of the agents is represented by 
a directed graph (digraph) 
${\cal{G}} = \left({\cal{V}}, {\cal{E}}\right)$, e.g., as defined in~\cite{Olfati_Murray_07}. 
Here, the  agents are represented by nodes $ {\cal{V}}= \left\{ 1, 2, \ldots, {n\!+\!1} \right\}$, $n>1$ and their connectivity by edges $ {\cal{E}}   \subseteq {\cal{V}} \times {\cal{V}} $, where each agent $j$ belonging to the set of 
neighbors $N_i  \subseteq {\cal{V}} $  of the agent $i$ satisfies  $ j \ne i$ and $(j,i) \in {\cal{E}} $.

\vspace{0.1in}
The evolution of the  multi-agent network is defined using the graph ${\cal{G}}$, as in Eq.~\eqref{system_eq}.  
The elements
 $l_{i,j}$ of the $(n+1)\times(n+1)$ Laplacian $L$  of the graph ${\cal{G}}$ are real and given by 
\begin{eqnarray}
\label{eq_Laplacian_defn}
l_{i,j} & =   \left\{ 
\begin{array}{ll}
-a_{i,j} < 0, 	& {\mbox{if}} ~ j \in N_i \\
\sum_{m=1}^{n+1} a_{i,m}, & {\mbox{if}} ~ j = i,  \\
0 & {\mbox{otherwise,}}
\end{array}  
\right.
\end{eqnarray}
where 
the weight $a_{i,j}$ is nonzero (and positive) 
if and only if $j$ is in the set of neighbors $N_i  \subseteq {\cal{V}} $  of the agent $i$, 
each row of the Laplacian $L$ adds to zero, i.e., from Eq.~\eqref{eq_Laplacian_defn}, 
the $(n+1) \times 1$ vector of ones  ${\textbf{1}}_{n+1}= [1, \ldots, 1]^T$  is a right eigenvector of the Laplacian $L$ with eigenvalue $0$,   
\begin{eqnarray}
\label{eq_first_lambda_eigenvector}
L{\textbf{1}}_{n+1} & = 0  {\textbf{1}}_{n+1} .
\end{eqnarray}

\subsubsection{Network dynamics}
One of the agents is assumed to be a virtual source agent \cite{leonard2001virtual}, which can be 
used to specify a desired consensus value $X_s$.  Without loss of 
generality, the state $\hat{X}_{n+1}$ of last $n+1$ node is assumed to be a virtual source agent $X_s$, where $s=n+1$. Moreover, each 
agent in the network should have access to the virtual source agent ${X}_s$  through the network, as formalized below. 
Note that this is a less stringent requirement than the graph without the virtual source being strongly connected. 

\vspace{0.1in}
\begin{Assum}[Rooted graph]
\label{assum_digraph_properties}
The digraph ${\cal{G}}$ is assumed  to have a directed path from the source node ${n+1}$ to any other  node $i$ in the graph, i.e.,  $ i \in {\cal{V}} \setminus \!{(n+1)}$. 
\qed 
\end{Assum}

\vspace{0.1in}
Some properties of the  graph  ${\cal{G}}$ without the source node $s=n+1$, i.e., ${\cal{G}}\!\setminus\!s$, 
are listed below.  
In particular, consider the  $n \times n$ pinned Laplacian  matrix $K$ associated with ${\cal{G}}\!\setminus\!s$ 
obtained by removing the row and column associated with the source node $n+1$ through  the partitioning of the Laplacian $L$, i.e., 
\begin{equation}
L  = 
\left[
\begin{array}{c}
\begin{array}{c|c}
K  &  -B  
\end{array}
\\[0.01in]
\hline 
\vspace{0.001in}
 L_b\\
\end{array}
\right]
\label{eq_def_pinned_K}
\end{equation}
where $L_b$
is the $1 \times (n+1)$ size row vector of Laplacian $L$ corresponding the source node $s=n+1$ and
$B$ is an  $n \times 1$ vector 
\begin{equation}
\label{eq_B_def}
\begin{array}{rl}
B & = [ a_{1,s}, a_{2,s}, \ldots, a_{n,s}]^T
 \\ &  
 ~= [ B_{1}, B_{2}, \ldots, B_{n}]^T , 
\end{array}
\end{equation}
and non-zero value of $B_j$ implies that the agent $j$ is 
directly connected to the source $X_s$. The properties of the pinned Laplacian $K$ follow from Assumption~\ref{assum_digraph_properties}, e.g., see~\cite{Olfati_Murray_07}.
\begin{enumerate}
\item 
The  pinned Laplacian matrix $K$  is invertible, i.e., 
 \begin{equation}
\det{(K)} \ne 0.
\label{eq_K_eigenvector}
 \end{equation}
 \item 
 The eigenvalues of the pinned Laplacian $K$ have strictly-positive, real parts.
 \item 
 The product of the inverse of the pinned Laplacian $K$ with $B$ leads to an  $n \times 1$ vector of  ones, ${\textbf{1}}_n$, i.e., 
\begin{eqnarray}
\label{eq_KinvtimesB}
K ^{-1} B & = {\textbf{1}}_n.
\end{eqnarray}
\end{enumerate}

\noindent 
The dynamics of the $n$ non-source agents with state vector, 
$X$ represented by the remaining graph ${\cal{G}}\!\setminus\!s$, can be  given by 
\begin{equation}
\begin{array}{rl}
X[k+1] &  = X[k]  -\alpha  K X[k] + \alpha   B  X_s[k]
 \\
&  = \left(   {\bf{I}_n} - \alpha  K     \right) X[k]   + \alpha   B  X_s [k] \\
& = P  X[k] + \alpha   B  X_s[k]  .
\label{system_non_source}
\end{array}
\end{equation}
where the matrix $P   ~ =  \textbf{I}_{n}-\alpha  K$, $ {\textbf{I}}_{n}$ is the $n\times n$ identity matrix, 
and $\alpha$ is the  update gain.

\vspace{0.1in}
\subsubsection{Stability conditions}
\label{Section_noDSR_stability}
Bounds can be established on the 
 update gain $\alpha$ to ensure stability. 
For  any  eigenvalue  $\lambda_{K,m} = a_m + j b_m$ of graph Laplacian $K$, with real part $a_m>0$ (from Assumption~\ref{assum_digraph_properties}) and imaginary part $b_m$, the corresponding eigenvalue of the matrix $P$ is
given by 
\begin{equation}
    \lambda_{P,m} = 1-\alpha(a_m + j b_m). 
\end{equation}
For stability of the non-source dynamics in Eq.~\eqref{system_non_source}, the magnitude of $\lambda_{P,m}$ needs to be less than one, i.e., 
\begin{equation}
    \left| 1-\alpha (a_m + j b_m)\right| <1, \forall\; m \in \{1,2, \hdots, n\}.
\end{equation}
This condition for stability is met if the update gain $\alpha$ satisfies~\cite{Devasia_2019_IJC}
\begin{equation}
    0 < \alpha < \min_{1 \leq m \leq n} \frac{2 a_m}{a_m^2 + b_m^2} = \overline{\alpha}.
     \label{eq_stability_condition_lem_Stability_and_Update_gain}
\end{equation}

\vspace{0.1in}
\subsubsection{Convergence to consensus}
\label{sec_Stable_consensus}
With a stabilizing  update gain $\alpha$ as in Eq.~\eqref{eq_stability_condition_lem_Stability_and_Update_gain}, the state $X$ of the network (of all non-source agents) converges to a fixed source value $X_s$, e.g., 
for a step change in the source value $X_s$ from $x_i$ to $x_f$, i.e., $X_s[k] = x_i$, $\forall\; k<0$ (initial desired state) and $X_s[k] = x_f$, $\forall\; k \ge 0$.
Since the eigenvalues $\lambda_{P,m}$  of the matrix
$P$ are inside the unit circle, the solution to Eq.~\eqref{system_non_source} for the step input 
converges,
\begin{equation}
    \begin{array}{rcl}
  X[k+1] - X[k]  &  = & 
  P \left( X[k] - X[k-1]  \right) \\
  & = & P^k \left( X[1] - X[0]  \right)
\rightarrow 0 ,
\end{array}
\label{Eq_controlled_gen_soln}
\end{equation}
as $ k \rightarrow \infty$ because $ P^{k} \rightarrow 0$. 
Thus, $\lim_{k\rightarrow\infty}   X[k+1] =\lim_{k\rightarrow\infty}   X[k] $, and from the first line of Eq.~\eqref{system_non_source}, 
\begin{equation}
  \lim_{k\rightarrow\infty} K X[k] =   B  x_f  .
\end{equation}  
As a result, from the invertibility of $K$ in Eq.~\eqref{eq_K_eigenvector}, 
and $K ^{-1} B  = {\textbf{1}}_n$ from Eq.~\eqref{eq_KinvtimesB}, the limit for the state $X[k]$ is found to be 
\begin{eqnarray}
X[k]  \rightarrow K^{-1} B  x_f 
= 
\rightarrow {\textbf{1}}_{n} x_f ~~\mbox{as}\quad k \rightarrow \infty.
\label{system_non_source_stability}
\end{eqnarray}
\noindent Thus, the control law in Eq.~\eqref{system_non_source} achieves consensus.

\vspace{0.1in}
\subsubsection{Spectral radius and rate of convergence}
The rate of convergence to consensus depends on the spectral radius $\sigma(P)$ of the matrix $P$ given by 
\begin{equation}
\sigma(P) =  \max_m | \lambda_{P,m} |  = \max_m | 1- \alpha \lambda_{K,m} | {<1}.
\label{Eq_matrix_norm_ineq_0}
\end{equation}
Note that for any $\epsilon > 0$, say 
\begin{equation}
    \epsilon = \frac{1-\sigma(P)}{2} > 0
\end{equation}
there exists a nonsingular matrix $Q$ such that the 
modified vector norm $\| X \| = \|Q X \|_\infty $  with the corresponding induced matrix norm  $\| \cdot \|$ satisfies, see~\cite{Ortega_87} (Section 5.3.5), 
\begin{equation}
\| P \| ~  \le  ~\sigma(P) + \epsilon 
~ = \frac{1 + \sigma(P) }{2} ~ < 1.
\label{Eq_matrix_norm_ineq}
\end{equation}
Hence, from Eq.~\eqref{Eq_controlled_gen_soln}, 
\begin{equation}
\begin{array}{rcl}
  \| X[k+1] - X[k] \|   &  \le  &  \| P \|^{k+1} \|   \textbf{1}_n (x_f- x_i)  \|   \\ 
  &  \le  & \left[ \sigma(P)+\epsilon \right]^{k+1} 
  \|   \textbf{1}_n ( x_f-   x_i )   \|.  
\end{array}
\label{Eq_rate_convergence_rate}
\end{equation}
Since $\epsilon$ can be chosen to be arbitrarily small, minimizing the spectral radius $\sigma(P)$ of the matrix $P$ 
results in faster convergence.

\vspace{0.1in}
\subsection{ {Convergence with structural robustness}} 
\label{Limit_of_convergence_rate}
The structural  robustness of the network's stability depends on the spectral radius $\sigma(P)$ of the matrix $P$ \cite{fagnani2017introduction}. 
For the network to be stable, the eigenvalues of the matrix $P$ need to be inside the unit circle. Hence,  
the smallest distance $d$ of its eigenvalues
$\lambda_{P,m} $ from the unit circle is a measure of the network's structural stability, i.e., robustness to perturbations, where 
\begin{equation}
d  =   1 - \sigma(P). 
\label{eq_optimal_no_dsr_robustness}
\end{equation}
Minimizing the spectral radius $\sigma(P)$ results in increased structural robustness.  
{ Therefore, rapid structurally-robust convergence} is achieved during transitions if the spectral radius $\sigma(P)$ is minimized. 
\noindent The optimal update gain $\alpha^*$ for minimum spectral radius $\sigma^*$, with the standard consensus dynamics (in Eq.~\eqref{system_non_source}) referred to as no-DSR approach hereon, can be found through a search based method, as

\begin{equation}
    \sigma^* = \min_{\alpha} \sigma(P) = \min_{0<\alpha <\overline{\alpha}} \left[   \max_m | 1- \alpha \lambda_{K,m} |\right].
\label{Eq_optimal_sigma_cmplx}
\end{equation}

\begin{Rem}{[Optimal no-DSR for real spectrum]}
\label{alpha_for_min_spectralradius_lemma}
For the special case when the graph has real spectrum, i.e., 
eigenvalues of the pinned Laplacian $K$ 
are real and satisfy
     \begin{equation} 
 0< \underline\lambda = \lambda_{K,1}  \le \lambda_{K,2}  \le \hdots \le  \lambda_{K,n} = \overline\lambda = \sigma(K) , 
 \label{eq_ordering_eigenvalues}
 \end{equation}
  the stability condition in Eq.~\eqref{eq_stability_condition_lem_Stability_and_Update_gain}, becomes 
\begin{equation}
    0<\alpha<  
     \frac{2}{ \left(     \overline \lambda \right)
      } = \overline{\alpha} .
    \label{Eq_noDSR_stablitity_limit_real}
\end{equation}
 If the extremal eigenvalues are distinct, i.e., 
 $\underline\lambda \ne \overline\lambda$, then 
 the update gain $\alpha^*$ that minimizes the spectral radius $ \sigma(P) $ is given by~\cite{xiao2004fast}
\begin{equation}
    \alpha^* = \frac{2}{\overline \lambda + \underline \lambda} <  
     \frac{2}{  \left(     \overline \lambda \right) }  , 
    \label{update_gain_for_max_robust}
\end{equation}
and 
the associated minimum spectral radius  is 
\begin{equation}
  \sigma^* =   \sigma(P^*) = \frac{\overline\lambda-\underline\lambda}{\overline\lambda+\underline\lambda}.
    \label{minimum_spectral_radius}
\end{equation}
If the extremal eigenvalues are the same, $\underline\lambda = \overline\lambda$ (e.g., in first-order platoon networks), then 
the spectral radius of the matrix ($\sigma(P)$) can be made the ideal value of zero, $ \sigma^*  =0$,  resulting in maximally fast convergence. 
 \end{Rem}

\subsection{The robust convergence optimization problem}
\vspace{-0.01in}
The range of acceptable update gain $\alpha$ in Eq.~\eqref{eq_stability_condition_lem_Stability_and_Update_gain}, limits the convergence rate. 
The research problem addressed is to  further reduce the spectral radius of the matrix $P$,  i.e. to improve 
the structural robustness and convergence rate, 
when each agent can modify  its update law 
\begin{enumerate}
\item 
using only existing information from the network neighbors, { and} 
\item 
without changing the network structure (network connectivity $K$).
\end{enumerate}

\vspace{0.1in}
\section{Proposed Solution}
\label{proposed_approach}
This section introduces the proposed  Accelerated Delayed Self Reinforcement (A-DSR) approach to 
achieve {
structurally-robust convergence} and establishes stability conditions.

\subsection{The A-DSR approach} 
\label{Section_General_ADSR_approach}
\vspace{0.1in}
\subsubsection{Graph's Laplacian potential}
\label{Sec_Graphs_Laplacian_potential}
For undirected graphs, the control law $u$ in  Eq.~\eqref{system_eq} can be considered as a  gradient-based search on the 
graph's Laplacian potential $\Phi_{{\cal{G}}}$
\cite{OlfatiSaberIEEETAC_04,Hongwei_lewis_12}
\begin{align}
\Phi_{{\cal{G}}}(\hat{X}) & =   \frac{1}{2} \sum_{i,j =1}^{n} a_{i,j}\left( \hat{X}_j - \hat{X}_i \right)^2
 = \hat{X}^T L \hat{X}, 
\label{system_eq_potential}
\end{align}
\noindent which results in the standard graph-based update law as in   Eq.~\eqref{system_eq}, 
\begin{align}
\hat{X}[k+1] & =   \hat{X}[k]   -   {\frac{\alpha}{2}} \nabla \Phi_{{\cal{G}}}(\hat{X}[k]) \nonumber \\
& =   \hat{X}[k] -  \alpha  L \hat{X}[k] .
\label{system_gradient}
\end{align}

\vspace{0.1in}
\subsubsection{Nesterov's accelerated-gradient-based update}
In general, the convergence of the gradient-based approach as in Eq.~\eqref{system_gradient} 
 can be improved using accelerated methods. In particular, applying  
the  Nesterov modification~\cite{Rumelhart_86,QIAN1999145} of the traditional gradient-based method to  
Eq.~\eqref{system_gradient}  
results in the accelerated-gradient-based modification of the system in Eq.~\eqref{system_eq}  to 
\begin{align}
\hat{X}[k+1] & = \hat{X}[k+1]
+ \beta \left(\hat{X}[k] -\hat{X}[k-1]  \right) 
\nonumber \\ 
 & \qquad 
-{\frac{\alpha}{2}}\nabla \Phi_{{\cal{G}}}\left\{\hat{X}[k] +\beta \left(\hat{X}[k] -\hat{X}[k-1]  \right)  \right\}  \nonumber \\ 
& =  \hat{X}[k] 
+  \beta\left( \hat{X}[k]-\hat{X}[k-1] \right)
\nonumber \\
& \qquad
-  \hat{\alpha}  L\left\{ \hat{X}[k] + \beta\left(\hat{X}[k]-\hat{X}[k-1] \right) \right\}    ,
\label{System_Nesterov_gradient_approach}
\end{align}
\noindent 
where $\beta$ is a scalar gain on the Nesterov-based terms and 
\begin{equation}
    \hat{\alpha} = \alpha(1+\beta).
    \label{Eq_alpha_hat_nesterov}
\end{equation}
Consequently,  the dynamics of the non-source agents 
$X$ represented by the remaining graph ${\cal{G}}\!\setminus\!s$, i.e., Eq.~\eqref{system_non_source}, becomes
\begin{equation}
\begin{array}{rl}
X[k+1] &  = X[k]   -\hat{\alpha}  K \left\{ X[k]  + \beta\left({X}[k]-{X}[k-1]  \right)
\right\} \\
&    \qquad + \beta\left({X}[k]-{X}[k-1] \right) \\
& \qquad+ \hat{\alpha}  B \{X_s[k] + \beta\left(X_s[k]-X_s[k-1]\right) \}
\end{array}
\label{accelerated_system_non_source}
\end{equation}

\noindent
The additional third term $ \beta \left({X}[k]-{X}[k-1] \right)$  
on the right hand side of Eq.~\eqref{accelerated_system_non_source} is referred to as the momentum term (this term alone forms the Heavy ball method in \cite{ghadimi2015global}) and  the similar terms 
 inside the curly   brackets of the second and fourth terms 
are referred to as the outdated-feedback addition.  

\vspace{0.1in}
\subsubsection{Directed graphs}
\noindent 
For general directed graphs, the  potential function $\Phi_{{\cal{G}}}$
in Eq.~\eqref{system_eq_potential} 
does not lead to the standard update equations~\cite{OlfatiSaberIEEETAC_04,Zhang_hui_IJC_2015}. 
Nevertheless, motivated by the gradient-based approach,  for each non-source agent, $1\le i \le n$, a modified potential can be considered as

\begin{align}
\Phi_{{\cal{G}},i} (\hat{X}) &  =   \sum_{j =1}^{n+1}  a_{i,j}\left(\hat{X}_i -\hat{X}_j \right)^2 . 
\label{system_eq_potential_local}
\end{align}

\noindent 
Here 
$\Phi_{{\cal{G}},i}$ is  a localized version of the graph's  Laplacian potential~\cite{OlfatiSaberIEEETAC_04,Zhang_hui_IJC_2015}, whose gradient with respect to $\hat{X}_i = X_i$

\begin{equation}
    u_i(\hat{X}) = - \frac{\alpha}{2} \frac{\partial \Phi_{{\cal{G}},i}}{\partial \hat{X}_i} = -\alpha K_i X + \alpha B_i X_s
    \label{single_agent_input}
\end{equation}

\noindent
with $K_i$ as the $i^{th}$ row of $K$, $B_i$ the $i^{th}$ row of the source connectivity vector $B$, will lead to the standard update equations for each agent's state $X_i$ in the state vector $X$ of non-source agents, as

\begin{equation}
    \begin{aligned}
    X_i[k+1] &= X_i[k] - \alpha K_i X[k] +\alpha B_i X_s[k].
    \end{aligned}
    \label{single_agent_gradient}
\end{equation}

\suppressfloats
\begin{figure}[!ht]
\begin{center}
\includegraphics[width=.90\columnwidth]{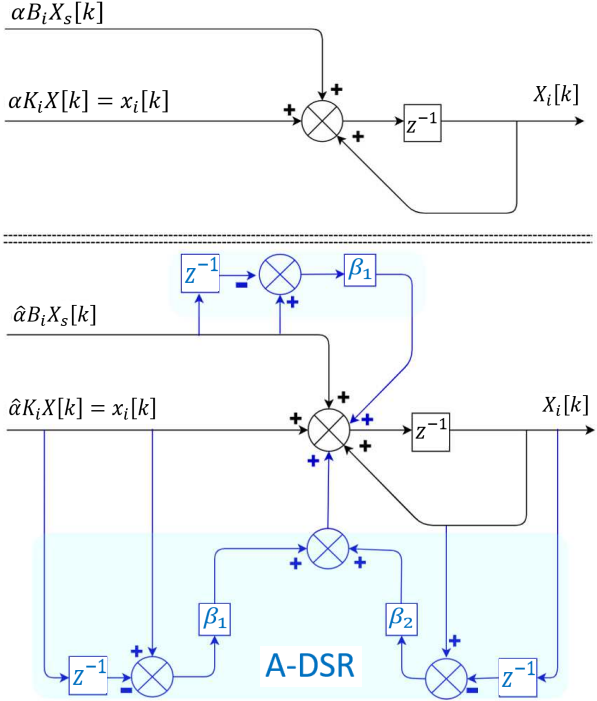}
\vspace{0.01in}
\caption{(Top) Implementation of standard consensus-based approach to multi-agent networks for the  $i^{th}$  agent as in Eq.~\eqref{system_non_source}. 
(Bottom) Accelerated delayed self reinforcement (A-DSR) approach for the $i^{th}$ agent  in Eq.~\eqref{accelerated_system_non_source_A_DSR}
without using additional network information.
}
\label{fig_1_control_implementation}
\end{center}
\end{figure}

\noindent 
The application of the  accelerated-gradient approach
for local potential $\Phi_{{\cal{G}},i} (\hat{X})$ in Eq.~\eqref{system_eq_potential_local} (which does not necessarily decrease the graph potential ($\Phi_{{\cal{G}}}(\hat{X})$) in Eq.~\eqref{system_eq_potential},~\cite{Rumelhart_86,QIAN1999145}) 
leads to the same Eq.~\eqref{accelerated_system_non_source}.

\vspace{0.1in}
\subsubsection{
{A-DSR update}}
\label{Proposed A-DSR}
The  Nesterov-update law in Eq.~\eqref{accelerated_system_non_source} uses the
same  gain $\beta$
for the momentum  and the outdated-feedback terms (Nesterov's accelerated method in \cite{Nesterov:2014:ILC:2670022}). A generalization of this is to use different gains $\beta_1, \beta_2$ for the outdated-feedback and momentum terms (respectively), as used before in optimization theory  \cite{lessard2016analysis},   
\begin{equation}
\begin{array}{rl}
X[k+1] &  = X[k]   -\hat{\alpha}  K \{ X[k]  + \beta_1\left({X}[k]-{X}[k-1] \right) \}   \\
&    \qquad + \beta_2\left({X}[k]-{X}[k-1] \right)\\
& \qquad+ \hat{\alpha}   B  \{X_s[k] + \beta_1\left(X_s[k]-X_s[k-1]\right) \}    .
\end{array}
\label{accelerated_system_non_source_A_DSR}
\end{equation}
\noindent 
where, from Eq.~\eqref{Eq_alpha_hat_nesterov}, $\hat{\alpha}=\alpha(1+\beta_1)$. The above  accelerated approach, is referred to as the accelerated delayed self reinforcement (A-DSR) in the following, since it does not require  additional information from the network, or having to change the network connectivity. Rather, each agent uses delayed versions of known information to reinforce its own update. 
To illustrate, for each non-source agent $i$, let $x_i$ be the information obtained from the network, i.e., 
\begin{equation}
\begin{array}{rl}
x_i[k]  &  =  \hat{\alpha} K_i  X[k] ,
\end{array}
\label{accelerated_system_non_source_2}
\end{equation}
where $K_i$ is  the $i^{th}$ row of the pinned Laplacian $K$.
Then, the update  of agent $X_i$ is,  from Eq.~\eqref{accelerated_system_non_source_A_DSR}, 
\begin{equation}
\begin{array}{rl}
X_i[k+1] 
& = X_i[k] - \{x_i[k]  +\beta_1(x_i[k] -x_i[k-1]) \} \\
&  \qquad + \beta_2\left({X}_i[k]-{X}_i[k-1] \right) \\
&\qquad+ \hat{\alpha}   B_i  \{X_s[k] + \beta_1\left(X_s[k]-X_s[k-1]\right) \} ,  
\end{array}
\label{accelerated_system_non_source_single}
\end{equation}
where  $B_i$ is the $i^{th}$ row of the source connectivity vector 
$B$. 
The delayed self-reinforcement (DSR) approach, however, requires each agent to store delayed versions $X_i[k-1]$ and $ x_i[k-1]$ of its current  state $X_i[k]$ and information $x_i[k]$ from the network, 
as illustrated in Fig.~\ref{fig_1_control_implementation}.  ~ \hfill \qed

\begin{Rem}
\label{different_update_methods}
The A-DSR method in Eq.~\eqref{accelerated_system_non_source_A_DSR} without the momentum term ( i.e., $\beta_2=0$) is referred to as the Outdated-feedback method,  
without the outdated-feedback term ( i.e., $\beta_1=0$) is referred to as the Momentum method, and with equal parameters 
( i.e., $\beta_1= \beta_2 = \beta$) is referred to as the Nesterov-update method.  
\end{Rem}

\begin{Rem}[Stability for directed graphs] The local Laplacian potential in Eq.~\eqref{system_eq_potential_local} whose gradient is used for deriving each agent's standard update (in Eq.~\eqref{single_agent_gradient}) and A-DSR approach (in Eq.~\eqref{accelerated_system_non_source_single}), doesn't reduce the overall Laplacian potential \cite{Zhang_hui_IJC_2015} of the directed graph. Thus the convergence studies from optimization theory (which require the graphs to be strongly connected, 
e.g.,~\cite{Makhdoumi_2015,Khan_Xin_2020}) cannot be used to establish stability for general directed graphs.
\end{Rem}

\subsection{
{ Stability of A-DSR}}
\label{Section_stability_proofs}
The stability conditions  for the   general A-DSR approach in Eq.~\eqref{accelerated_system_non_source_A_DSR}  are presented below.

\subsubsection{Diagonalizing the pinned Laplacian}
\label{section_diagonalization_stability}
The network with A-DSR in Eq.~\eqref{accelerated_system_non_source_A_DSR}  can be  decomposed into subsystems using an 
invertible transformation matrix $P_K$
as 
\begin{equation}
    \begin{aligned}
    X[k] = P_K X_J[k], 
    \end{aligned}
    \label{velocity_transform_J}
\end{equation}
where the 
transformation matrix $P_K$ is selected to diagonalize the  pinned Laplacian $K$ as
\begin{equation}
    \begin{aligned}
    K_J = P_K^{-1} K P_K
    \end{aligned}
    \label{jordan_form_K}
\end{equation}
where the diagonal terms of matrix $K_J$ are the eigenvalues $ \lambda_{K,m}$ for $m= {1,2, ..., n}$, which can be complex and with multiplicity greater than 1. 
Since input doesn't affect stability, setting $X_s[k] = 0, \; \forall\, k$, and pre-multiplying the Eq.~\eqref{accelerated_system_non_source_A_DSR} with $P_k^{-1}$ results in  

\begin{equation}
    \begin{aligned}
    &X_J[k+1] - X_J[k]  +\hat{\alpha} K_J(X_J[k]\\ &+ \beta_1(X_J[k]-X_J[k-1]))\\ 
    & - \beta_2 (X_J[k]-X_J[k-1]) = 0. 
    \end{aligned}
    \label{general_nesterov_jordan_form}
\end{equation}
\noindent 
The stability of network with A-DSR in Eq.~\eqref{accelerated_system_non_source_A_DSR} is equivalent to the stability of Eq.~\eqref{general_nesterov_jordan_form} in the transformed coordinate.

\vspace{0.1in}
\subsubsection{Characteristic equations}
Taking the z-transform of Eq.~\eqref{general_nesterov_jordan_form} results in 
 \begin{equation}
     \begin{aligned}
     &(z^2 {\textbf{I}}_{n} - z \left[ (1+\beta_2){\textbf{I}}_{n} - \hat{\alpha} (1+\beta_1) K_J \right] 
     \\ &
     - \left(\hat{\alpha} \beta_1 K_J  - \beta_2 {\textbf{I}}_{n} \right) ) X_J(z) = 0. 
     \end{aligned}
     \label{general_z_transform}
 \end{equation}
Therefore, the network with A-DSR update in Eq.~\eqref{accelerated_system_non_source_A_DSR} is stable if and only if, for each eigenvalue $\lambda_{K,m}$ of the pinned Laplacian $K$,  the roots of the following characteristic equation  
 \begin{equation}
     \begin{aligned}
     D(z) & = z^2
     +z \left[  \hat{\alpha} (1+\beta_1){ \lambda_{K,m}   -(1+\beta_2)   }  \right]  \\
     & \quad \quad  + (\beta_2 -\hat{\alpha}\beta_1 { \lambda_{K,m}     } )
     \quad = 0 
     \end{aligned}
     \label{general_char_eq}
 \end{equation} 
have magnitude less than one. 
For the case of complex eigenvalue $\lambda_{K,m} = a_m + j b_m$, the 
real Jordan form of the z-transform of the diagonalized general A-DSR update equation in Eq.~\eqref{general_nesterov_jordan_form}, for the block associated with the Laplacian eigenvalue  pair $a_m\pm jb_m$ is
\begin{equation}
    \begin{aligned}
     z^2 {\textbf{I}}_{2} -&z \left((1+\beta_2){\textbf{I}}_{2}
     -\hat{\alpha}(1+\beta_1)\begin{bmatrix}
a_m & b_m \\
-b_m & a_m
\end{bmatrix}\right) \\
-& \left( \hat{\alpha} \beta_1\begin{bmatrix}
a_m & b_m \\
-b_m & a_m
\end{bmatrix} - \beta_2 {\textbf{I}}_{2} \right), 
    \end{aligned}
    \label{complex_char_matrix_eq}
\end{equation}
where ${\textbf{I}}_{2}$ denotes an identity matrix of size $2 \times 2$, $a_m$ and $b_m$ are the real and imaginary parts of $\lambda_{K,m}$. The determinant of Eq.~\eqref{complex_char_matrix_eq} yields a fourth order equation of the form 

\begin{equation}
    D(z) =  z^4 + a_3 z^3 + a_2 z^2 + a_1 z + a_0 = 0, 
    \label{complex_char_eq_complex}
\end{equation}

\noindent where 
\begin{equation}
    \begin{aligned}
    a_0 &= (a_m \hat{\alpha} \beta_1 - \beta_2)^2 + \hat{\alpha}^2 b_m^2 \beta_1^2, \\
    a_1 &= -2\{ (a_m \hat{\alpha} \beta_1 - \beta_2)^2 + a_m \hat{\alpha} \beta_1(a_m \hat{\alpha} - 1) \\ &~+ \hat{\alpha}  (\hat{\alpha} b_m^2 \beta_1 - {a_m}\beta_2) + \hat{\alpha}^2 b_m^2\beta_1^2 + \beta_2  \}, \\
    a_2 &= (a_m \hat{\alpha} \beta_1 - \beta_2)^2 + (a_m \hat{\alpha} - 1)^2 + 2 \hat{\alpha}^2 \beta_1(a_m^2+b_m^2), \\
    &~ + 4 \beta_2 -2 a_m \hat{\alpha} (2\beta_1+\beta_2) + \hat{\alpha}^2 b_m^2 (\beta_1^2+1), \\
    a_3 &= 2 a_m \hat{\alpha} (\beta_1+1) - 2 (\beta_2+1) .
    \end{aligned}
    \label{lemma2_a0toa4_coeff_Def}
\end{equation}
Hence, for the complex eigenvalue case, the stability of the  A-DSR (in Eq.~\eqref{accelerated_system_non_source_A_DSR}) can be determined by obtaining conditions for the roots of Eq.~\eqref{complex_char_eq_complex} to be within unit circle on the complex plane.

\vspace{0.1in}
\subsubsection{Stability conditions}
Stability conditions follow from the Jury test.
\vspace{0.1in}
\begin{Lem}{[Jury test based stability]}
\label{general_ADSR_lemma_complex}
The generalized A-DSR in Eq.~\eqref{accelerated_system_non_source_A_DSR} is stable if and only if the A-DSR gains $\hat{\alpha},\; \beta_1$ and $\beta_2$ satisfy the following conditions, for  each eigenvalue $\lambda_{K,m}$ of the pinned Laplacian $K$.

\begin{enumerate} 

\item 
If the eigenvalue $\lambda_{K,m} = a_m $ is real valued, then 
\begin{equation}
    \begin{aligned}
     (i)~ &
    0  < \hat{\alpha}  \\ 
      (ii)~ &
    \left[ \hat{\alpha} { a_m     }(\beta_1 +\frac{1}{2})-1 \right] < \beta_2 < \left( \hat{\alpha} \beta_1 
    { a_m
    }+1 \right)  
    \end{aligned}
    \label{general_stability_cond}
\end{equation}

\item 
If the eigenvalue $\lambda_{K,m} = a_m + j b_m$ is complex valued (i.e., $b_m \ne 0$), then 
\begin{equation}
    \begin{aligned}
    (i)~ & 0 < \hat{\alpha}^2 ,   \\ 
   (ii)~ &  0 < (2(\beta_2+1)-\hat{\alpha}(2\beta_1+1)a_m)^2  \\
   & \qquad \qquad + \hat{\alpha}^2(2\beta_1+1)^2b_m^2 , \\
    (iii)~ & -1 <  (a_m \hat{\alpha} \beta_1 - \beta_2)^2 + \hat{\alpha}^2 b_m^2 \beta_1^2 < 1, \\ 
    (iv)~ & |a_0a_3-a_1|  < |a_0^2-1|,  \\ 
    (v)~ &  |a_2  (a_0^2-1)(a_0-1)  -(a_0a_1-a_3)(a_0a_3-a_1)| \\  
    & ~ 
    < |(a_0^2-1)^2-(a_0a_3-a_1)^2|.  
    \end{aligned}
    \label{stability_ADSR_complex}
\end{equation}

\end{enumerate}
\end{Lem}

\vspace{0.1in}

{\bf{Proof}~}
If the eigenvalue $\lambda_{K,m}$ is real valued, 
then, the Jury test leads to the following three  necessary and sufficient  conditions for
the roots of the characteristic equation
in Eq.~\eqref{general_char_eq} 
to have magnitude less than one.

\begin{enumerate}
    \item $D(z=1) > 0$
    \begin{equation}
    \begin{aligned}
        &1 + 
        \left[ \hat{\alpha}(1+\beta_1){\lambda_{K,m}     } -(1+\beta_2) \right] +(\beta_2-\hat{\alpha} \beta_1 { \lambda_{K,m}     }) >0 \\
        &=> \hat{\alpha} > 0, 
    \end{aligned}
    \label{eq_alpha_positive_condition}
    \end{equation}
    which is satisfied due to the first condition in Eq.~\eqref{general_stability_cond}. 
     \item $(-1)^2D(z=-1) > 0$
    \begin{equation}
    \begin{aligned}
       &1 - \left[ \hat{\alpha}(1+\beta_1){ \lambda_{K,m}     } -(1+\beta_2) \right] 
        +(\beta_2-\hat{\alpha} \beta_1 { \lambda_{K,m}     }) >0  \\
        &=> \beta_2 > 
        \hat{\alpha} { \lambda_{K,m}     } \left(\beta_1 +\frac{1}{2}\right)-1 
    \end{aligned}
    \end{equation}
    
    or 
        \begin{equation}
     (\hat{\alpha} \beta_1 { \lambda_{K,m}     }-1) + \frac{\hat{\alpha}  { \lambda_{K,m}     }}{2} < \beta_2.
     \label{Lemma_1_Proof_cond_2}
       \end{equation}
    
    \item 
     $ | D(z=0)|  < 1$ 
    $$|\beta_2 -\hat{\alpha}\beta_1 { \lambda_{K,m}     }|<1$$ or 
     \begin{equation}
    \begin{aligned}
        (\hat{\alpha} \beta_1 { \lambda_{K,m}     }-1) < \beta_2 < (1+\hat{\alpha} \beta_1 { \lambda_{K,m}     }).
    \end{aligned}
         \label{Lemma_1_Proof_cond_3}
    \end{equation}
\end{enumerate}

\noindent 
As $\hat{\alpha}  { \lambda_{K,m}     } > 0 $ (since $\hat{\alpha}>0$ from Eq.~\eqref{eq_alpha_positive_condition} and $ { \lambda_{K,m}     }>0$ from Assumption~\ref{assum_digraph_properties}), 
the condition in Eq.~\eqref{Lemma_1_Proof_cond_2}  is more stringent than the lower bound on $\beta_2$ in Eq.~\eqref{Lemma_1_Proof_cond_3}, resulting in condition (ii) of Eq.~\eqref{general_stability_cond}.

\vspace{0.1in}
 If the eigenvalue $\lambda_{K,m}=a_m + jb_m$ is complex valued ($b_m \neq 0$), 
then, the Jury test leads to the following necessary and sufficient conditions for
stable roots of the characteristic equation
in Eq.~\eqref{complex_char_eq_complex}.

\begin{enumerate}
    \item $D(z=1) > 0$
    \begin{equation}
    \begin{aligned}
        &1 + a_3 + a_2 + a_1 + a_0 > 0, \\
        &=> \hat{\alpha}^2 (a_m^2 + b_m^2 ) > 0,
    \end{aligned}
    \end{equation}
    \noindent which can be simplified further (since $a_m>0$ from Assumption~\ref{assum_digraph_properties}) as condition (i) in Eq.~\eqref{stability_ADSR_complex}.

    \item $(-1)^4 D(z=-1)>0$  resulting in condition~(ii) of Eq.~\eqref{stability_ADSR_complex} since 
      \begin{equation}
         \begin{aligned}
         & 0 < 1 - a_3 + a_2 - a_1 + a_0  \\
         &= (2(\beta_2+1)-\hat{\alpha}(2\beta_1+1)a_m)^2  \\
         & \qquad \qquad + \hat{\alpha}^2(2\beta_1+1)^2b_m^2 .
        \end{aligned}
    \end{equation}

    \item $| D(z=0) | < 1$
        \begin{equation}
            \begin{aligned}
             &| (a_m \hat{\alpha} \beta_1 - \beta_2)^2 + \hat{\alpha}^2 b_m^2 \beta_1^2 |<1.\\ 
             &=> -1 <  (a_m \hat{\alpha} \beta_1 - \beta_2)^2 + \hat{\alpha}^2 b_m^2 \beta_1^2 < 1. 
            \end{aligned}
        \end{equation}    
 \end{enumerate}
   In addition to the above three conditions (which are similar to the real eigenvalue case), the complex case has two additional stability conditions (iv) and (v) in Eq.~\eqref{stability_ADSR_complex} from the Jury test. 
\qed

\vspace{0.1in}
\subsubsection{Robust stability with general A-DSR}
Independently varying the gains of momentum and outdated-feedback terms gives additional flexibility, which can be used to further improve the robust  convergence when compared to the case without A-DSR. More formally, the general A-DSR approach can be used to  minimize the maximum magnitude $\bar{z}_m$ of the roots $(z_{\lambda_{K,m},1}, z_{\lambda_{K,m},2})$ of the characteristic equation $D(z)=0$ in Eq.~\eqref{general_char_eq} associated with the eigenvalues  $\lambda_{K,m}$ of the pinned Laplacian $K$, i.e., 
\begin{equation}
    \sigma^* = \min_{{\hat{\alpha}}, \beta_1, \beta_2}  \left[   \max_m ( \bar{z}_m  )  |\right], 
\label{Eq_optimal_sigma_cmplx}
\end{equation}
where 
 $\bar{z}_m = \max (|z_{\lambda_{K,m},1}|, |z_{\lambda_{K,m},2}|)$.

\vspace{0.1in}
\subsection{Graphs with  real spectrum}
\label{subsection_robust_convergence_with_ADSR}

 In general, using different gains for the momentum  and outdated-feedback terms (i.e., different values of $\beta_1, \beta_2$) can yield better performance than using the same gains for each term. However, for graphs with real spectrum (which includes all undirected graphs), the momentum term is sufficient to yield fast convergence and balanced robustness, as shown below. 

\vspace{0.1in}

\begin{Assum}[Real spectrum] In this section, the pinned Laplacian $K$ is assumed to have real eigenvalues, ordered as in Eq.~\eqref{eq_ordering_eigenvalues}.
\end{Assum}

\vspace{0.1in}
\subsubsection{Stability given range of Laplacian eigenvalues}
The application of 
Lemma~\ref{general_ADSR_lemma_complex}  requires knowledge of all eigenvalues $ {\lambda_{K,m}     }$  of the pinned Laplacian $K$. The following corollary provides sufficient conditions for stability in terms of the range $[\underline\lambda~ \overline\lambda]$ of the eigenvalues $ {\lambda_{K,m}}$ from Eq.~\eqref{eq_ordering_eigenvalues}. 
To begin, the  stability condition for general A-DSR update in Eq.~\eqref{general_stability_cond} is used to  deduce stability for the other (Nesterov-update, momentum and outdated-feedback defined in Remark~\ref{different_update_methods}) methods for graphs with real spectrum. 

\begin{Cor}
\label{corrolary_Acc_methods}
The network update as in Eq.~\eqref{accelerated_system_non_source_A_DSR}, for the following accelerated methods,
is stable if and only if $\hat{\alpha} > 0$, and the gains  satisfy the following {for each eigenvalue  $\lambda_{K,m}$ of the pinned Laplacian $K$. }
\begin{enumerate}
    \item Nesterov-update method in Eq.~\eqref{accelerated_system_non_source}\cite{Devasia_ICPS_2019} 
    with $\beta_1= \beta_2 = \beta$:
    \begin{equation}
        \begin{aligned}
       \frac{\hat{\alpha}{ \lambda_{K,m}     }}{2}-1 < \beta (1 - \hat{\alpha} { \lambda_{K,m}     }) < 1.
        \end{aligned}
        \label{nesterov_stability_1}
    \end{equation}
    
    \item Momentum method ($\beta_1 = 0$):
    \begin{equation}
        \begin{aligned}
        \frac{\hat{\alpha} {\lambda_{K,m}     }}{2}-1 < \beta_2 < 1.
        \end{aligned}
        \label{momentum_stability}
    \end{equation}
    
    \item Outdated-feedback method ($\beta_2 = 0$):
    \begin{equation}
        \begin{aligned}
        -1 &<  \hat{\alpha} { \lambda_{K,m}     }  \beta_1 < 1 - \frac{\hat{\alpha} { \lambda_{K,m}     }}{2}.
        \end{aligned}
        \label{acc_stability}
    \end{equation}
\end{enumerate}
\end{Cor}
\vspace{0.1in}
{\bf{Proof}~}
For the Nesterov-update method  ($\beta_1=\beta_2=\beta$), the stability condition in Eq.~\eqref{general_stability_cond} becomes
\begin{equation}
\begin{aligned}
     \left[ \hat{\alpha} { \lambda_{K,m}     }(\beta +\frac{1}{2})-1 \right] &< \beta < \left( \hat{\alpha} \beta 
    { \lambda_{K,m}
    }+1 \right), 
\end{aligned}
\label{eq_proof_cor_1_1}
\end{equation}
and  subtracting $\hat{\alpha} \beta\lambda_{K,m}$ from both sides results in  Eq.~\eqref{nesterov_stability_1}. 
For the momentum method, 
Eq.~\eqref{general_stability_cond} becomes  Eq.~\eqref{momentum_stability} with $\beta_1=0$.
For the outdated-feedback method, with $\beta_2=0$,  Eq.~\eqref{general_stability_cond} becomes
\begin{equation}
         \left[\hat{\alpha} { \lambda_{K,m}     }(\beta_1 +\frac{1}{2})-1 \right] ~< 0 < \left( \hat{\alpha} \beta_1 
    { \lambda_{K,m}
    }+1 \right).
    \label{corr_1_proof_ac_method_1_1}
\end{equation}
The left inequality in Eq.~\eqref{corr_1_proof_ac_method_1_1} can be simplified to
\begin{equation}
    \hat{\alpha}\lambda_{K,m} \beta_1< 1 - \frac{\hat{\alpha}\lambda_{K,m}}{2}
\end{equation}
and the right inequality becomes 
\begin{equation}
   \hat{\alpha} \lambda_{K,m} \beta_1 > -1, 
\end{equation}
resulting in the stability condition  in Eq.~\eqref{acc_stability}. 
\qed

\vspace{0.1in} 
\begin{Cor}
\label{corrolary_Acc_methods_range}
The network update as in Eq.~\eqref{accelerated_system_non_source_A_DSR}, for the following accelerated methods,
is stable if and only if $\hat{\alpha} > 0$, and the gains satisfy the following, where 
\begin{equation}
    \begin{aligned}
 \lambda_{*} & =
 \left\{ 
 \begin{array}{rcl} 
\underline{\lambda} 
& {\mbox{if}} & \beta_1 \le  -\frac{1}{2}  \\
\overline{\lambda} 
& {\mbox{if}} & \beta_1 >   -\frac{1}{2}
 \end{array}
 \right. 
 \\
  \lambda^{*} & =
 \left\{ 
 \begin{array}{rcl} 
\overline{\lambda} 
& {\mbox{if}} & \beta_1 \le  0  \\
\underline{\lambda} 
& {\mbox{if}} & \beta_1 >   0
 \end{array}
 \right.
    \end{aligned}
    \label{general_stability_cond_eigVal_range_2}
\end{equation}
\begin{enumerate} 
\item 
Generalized A-DSR method: 
\begin{equation}
    \begin{aligned}
    \left[ \hat{\alpha} { \lambda_{*}     }(\beta_1 +\frac{1}{2})-1 \right] &< \beta_2 < \left( \hat{\alpha} \beta_1 
    { \lambda^{*}
    }+1 \right) .  
    \end{aligned}
    \label{general_stability_cond_eigVal_range_1}
\end{equation}
\item Nesterov-update method  ($\beta_1=\beta_2=\beta$): 
\begin{equation}
\begin{aligned}
     \left[ \hat{\alpha} { \lambda_{*}     }(\beta +\frac{1}{2})-1 \right] &< \beta < \left( \hat{\alpha} \beta 
    { \lambda^*
    }+1 \right).
\end{aligned}
\label{eq_proof_cor_3_1}
\end{equation}
    \item Momentum method ($\beta_1 = 0$):
     \begin{equation}
        \begin{aligned}
        \frac{\hat{\alpha} \lambda_*}{2}-1 &< \beta_2 < 1.
       \end{aligned}
        \label{momentum_stability_extremum}
    \end{equation}
    \item Outdated-feedback method ($\beta_2 = 0$):
    \begin{equation}
         \left[ \hat{\alpha} { \lambda_*     }(\beta_1 +\frac{1}{2})-1 \right] ~< 0 < \left( \hat{\alpha} \beta_1 
    { \lambda^*
    }+1 \right).
    \label{corr_1_proof_ac_method_3_1}
\end{equation}

\end{enumerate}
\end{Cor}
\vspace{0.1in}

\vspace{0.1in}
{\bf{Proof}~}
This follows from Lemma~\ref{general_ADSR_lemma_complex} and the proof of  Corollary~\ref{corrolary_Acc_methods} since 
$$\hat{\alpha} \lambda_{K,m} (\beta_1 + \frac{1}{2} ) \le \hat{\alpha} \lambda_{*} \beta_1 , \quad  
 \hat{\alpha} \lambda^{*} \beta_1  \le \hat{\alpha} \lambda_{K,m} \beta_1 $$ 
 for all eigenvalues $\lambda_{K,m} $ of the pinned Laplacian $K$. Therefore, the conditions in this corollary are more stringent that the conditions in Lemma~\ref{general_ADSR_lemma_complex} and   Corollary~\ref{corrolary_Acc_methods}.
\qed

\subsubsection{Optimal A-DSR for graphs with real spectrum}
Fast convergence with structural robustness for A-DSR in networks with real spectrum is presented below, which is similar to the 
structurally-robust convergence without A-DSR in Section~\ref{Limit_of_convergence_rate}.
Note that the characteristic equation in  Eq.~\eqref{general_char_eq} with A-DSR for networks with real spectrum is equivalent to that of a standard  second order system of the form, 

\begin{equation}
    \begin{aligned}
    D(z) & = 
     z^2 + 2\zeta_{\lambda_{K,m}} \omega_{\lambda_{K,m}} z + \omega_{\lambda_{K,m}}^2 =0, 
    \end{aligned}
    \label{general_char_eq_normal_form}
\end{equation}

\noindent
where 
\begin{equation}
\begin{aligned}
\omega_{\lambda_{K,m}}^2 & =  (\beta_2 -\hat{\alpha}\beta_1 { \lambda_{K,m}     } ), \\ \zeta_{\lambda_{K,m}} & = \frac{ \hat{\alpha} (1+\beta_1) \lambda_{K,m}   -(1+\beta_2)     }{2 \omega_{\lambda_{K,m}}} , 
\end{aligned}
\label{Eq_def_damping_nat_freq}
\end{equation}

\noindent
with two roots
($ z_{\lambda_{K,m},i}$, $i \in \left\{1,2\right\})$ 
associated with each  real eigenvalue $\lambda_{K,m}$ of the pinned Laplacian $K$. As in the case without A-DSR, the goal is to select the roots 
($z_{\underline\lambda,i}, z_{\overline\lambda,i}, i\in\left\{1,2\right\}$) 
of the characteristic equation in Eq.~\eqref{general_char_eq} for A-DSR,  associated with the 
extremal eigenvalues $\lambda ={\underline{\lambda}, \overline{\lambda}}$ of the pinned Laplacian $K$, to be  equidistant from origin (for similar {structural robustness}) 
\begin{equation}
| z_{\underline\lambda}| = 
| z_{\underline\lambda,1}| = 
| z_{\underline\lambda,2}| = 
|z_{\overline\lambda,1}| =
|z_{\overline\lambda,2}| =
|z_{\overline\lambda}| 
 \label{Eq_ideal_ADSR_1}
\end{equation}
and be farthest away from the unit circle (for fast convergence), i.e., by choosing the A-DSR parameters $\hat{\alpha}, \beta_1, \beta_2$ to solve the following minimization problem
\begin{equation}
\min_{\hat{\alpha}, \beta_1, \beta_2} 
\left[ 
| z_{\underline\lambda}| = 
|z_{\overline\lambda}| 
\right].
 \label{Eq_ideal_ADSR_2}
\end{equation}
Furthermore,  the roots of  Eq.~\eqref{general_char_eq} associated with the 
dominant eigenvalue  $\underline\lambda$ of the pinned Laplacian are critically damped and positive, i.e., 
\begin{equation}
\zeta_{\underline\lambda} = -1, 
\quad  \quad 
 z_{\underline\lambda,1} = z_{\underline\lambda,2} > 0, 
 \label{Eq_ideal_ADSR_3}
\end{equation}
 as in the case without A-DSR, which can help to reduce  oscillations in the response.

\vspace{0.1in}
\begin{Lem}
{[Parameter selection for Robust A-DSR]}
\label{robust_ADSR_lemma}
{
Let (i)~the A-DSR parameter be chosen to be positive $\hat{\alpha}>0$ to meet the stability condition in Eq.~\eqref{general_stability_cond}, and (ii) the 
pinned Laplacian $K$ have at least two distinct eigenvalues, i.e., $\overline{\lambda}\neq\underline{\lambda}$ in 
Eq.~\eqref{eq_ordering_eigenvalues}.
Then, the A-DSR parameters  ($\hat{\alpha}, \beta_1, \beta_2$)
\begin{equation}
\begin{aligned}
     \hat{\alpha} &= \frac{4}{(\sqrt{\overline\lambda}+\sqrt{\underline\lambda})^2},~~  
    \beta_1 = 0, ~~
    \beta_2 = \frac{(\sqrt{\overline\lambda}-\sqrt{\underline\lambda})^2}{(\sqrt{\overline\lambda}+\sqrt{\underline\lambda})^2}
\end{aligned}
    \label{gains_lemma_for_robust_ADSR}
\end{equation}
result in 
\begin{enumerate}
    \item
    balanced robustness of the extremal modes, i.e., satisfies Eq.~\eqref{Eq_ideal_ADSR_1}, 
    the roots 
    $z_{\underline\lambda,i}, z_{\overline\lambda,i}, i\in\left\{1,2\right\}$ 
    as in Eq.~\eqref{Eq_ideal_ADSR_1}
    \item critical damping of the dominant mode, i.e., $\zeta_{\underline{\lambda}}=-1$ as in Eq.~\eqref{Eq_ideal_ADSR_3}, and 
    \item 
    optimal convergence, i.e., achieves the minimization in Eq.~\eqref{Eq_ideal_ADSR_2}.
\end{enumerate}
}
\end{Lem}

\vspace{0.1in}
\noindent
{\bf{Proof}~} 
This is shown below in four  steps. 

\vspace{0.05in}
\noindent 
{\bf{Step 1}} is to show that 
the roots  of  Eq.~\eqref{general_char_eq} associated with the 
extremal eigenvalue  $\overline\lambda$ of the pinned Laplacian cannot be overdamped. 
Note that if the damping ratio $\zeta_{\overline\lambda}$ of the 
 roots $z_{\overline\lambda,1}, z_{\overline\lambda,2}$ in   Eq.~\eqref{general_char_eq} associated with the 
extremal eigenvalue  $\overline\lambda$ 
is larger than one in magnitude, i.e.,  $| \zeta_{\overline\lambda}| > 1$, then the roots  
\begin{equation}
\begin{aligned}
z_{\overline\lambda,1} & =-\left(\zeta_{\overline\lambda} ~ \omega_{\overline\lambda}
\right) + 
\omega_{\overline\lambda}
\sqrt{\zeta_{\overline\lambda}^2 
-1} \\
z_{\overline\lambda,2} & =
-\left(\zeta_{\overline\lambda} ~ \omega_{\overline\lambda}
\right) - 
\omega_{\overline\lambda}
\sqrt{\zeta_{\overline\lambda}^2 -1}~
,  
\end{aligned}
\label{roots_of_2ndorder_Lemma3_1}
\end{equation}
are real and distinct and have 
different magnitudes 
$ |z_{\overline\lambda,1}| \ne 
|z_{\overline\lambda,2}| $, which cannot satisfy the lemma's equidistant condition as in Eq.~\eqref{Eq_ideal_ADSR_1}. 
Therefore, the roots $z_{\overline\lambda,1}, z_{\overline\lambda,2}$ of  Eq.~\eqref{general_char_eq} associated with the 
extremal eigenvalue  $\overline\lambda$ of the pinned Laplacian cannot be overdamped, i.e., \begin{equation}
 | \zeta_{\overline\lambda}| \le 1. \label{eq_damping_values_Lemma3_2}
\end{equation}

\vspace{0.05in}
\noindent 
{\bf{Step 2}} is to show that 
the equidistant condition of the lemma, as in Eq.~\eqref{Eq_ideal_ADSR_1}, leads to a zero outdated-feedback gain, $\beta_1=0$. 
Since the magnitude of the damping ratio is not more than one,  
$ | \zeta_{\overline\lambda}| \le 1$ from Eq.~\eqref{eq_damping_values_Lemma3_2},  the term $\zeta_{\overline\lambda}^2 
-1$ becomes non-positive in Eq.~\eqref{roots_of_2ndorder_Lemma3_1}, and therefore its square root is  either a complex number (when $|\zeta_{\overline{\lambda}}|<1$) or zero (when $|\zeta_{\overline{\lambda}}|=1$), and thus the magnitudes of the roots  become 
\begin{equation}
    |z_{\overline\lambda,1}| = |z_{\overline\lambda,2}| = |z_{\overline\lambda}| = \omega_{\overline\lambda}  = \sqrt{\beta_2-\hat{\alpha}\beta_1\overline\lambda} ~.
\end{equation}
Similarly, the magnitudes of the roots associated with the extremal value $\underline\lambda$
with damping ration $\zeta_{\underline\lambda} = -1 $ in Eq.~\eqref{Eq_ideal_ADSR_3}, are 
\begin{equation}
    |z_{\underline\lambda,1}| = |z_{\underline\lambda,2}| = |z_{\underline\lambda}| = \omega_{\underline\lambda}  = \sqrt{\beta_2-\hat{\alpha}\beta_1\underline\lambda} ~.
\end{equation}
To satisfy the equidistant condition, 
$$ |z_{\underline\lambda}| = \sqrt{\beta_2-\hat{\alpha}\beta_1\underline\lambda}
= 
\sqrt{\beta_2-\hat{\alpha}\beta_1\overline\lambda} = 
|z_{\overline\lambda}|,$$
and since 
$\hat{\alpha}>0$ and $\underline\lambda \ne \overline\lambda$,  $\beta_1=0$. Thus,
the magnitude of the roots (associated with the extremal eigenvalues) are 
\begin{equation}
    |z_{\overline\lambda}| = |z_{\underline\lambda}| =
    \omega_{\overline\lambda}  =
    \omega_{\underline\lambda}  =\sqrt{\beta_2}
    \label{eq_Lemma_3_proof_mag_beta1_0}
\end{equation}

\vspace{0.05in}
\noindent 
{\bf{Step 3}} is to show that 
the roots  of  Eq.~\eqref{general_char_eq} associated with the 
extremal eigenvalue  $\overline\lambda$ are critically damped. 
Using the damping ratio definition for the extremal modes, $\zeta_{\overline\lambda}$ and $\zeta_{\underline\lambda}$ in  Eq.~\eqref{Eq_def_damping_nat_freq},  with $\beta_1=0$ and $\zeta_{\underline\lambda}= -1$, and
substituting for $\omega_{\overline\lambda}, 
    \omega_{\underline\lambda}$ from Eq.~\eqref{eq_Lemma_3_proof_mag_beta1_0}, results in 
\begin{equation}
\begin{aligned}
    -1 &= \frac{\hat{\alpha}\underline\lambda-(1+\beta_2)}{2\sqrt{\beta_2}} \\
    \zeta_{\overline\lambda} &= \frac{\hat{\alpha}\overline\lambda-(1+\beta_2)}{2\sqrt{\beta_2}}
    \end{aligned}
    \label{Eq_extremal_mode_damping_exprsn}
\end{equation}

\noindent Solving the two equations in Eq.~\eqref{Eq_extremal_mode_damping_exprsn} for the magnitude $\sqrt{\beta_2}$ of the extremal roots results in 
\begin{equation}
    \sqrt{\beta_2} =\frac{ \hat{\alpha} (\overline\lambda-\underline\lambda)}{2(1+\zeta_{\overline\lambda})} , \label{Eq_extremal_mode_general_magnitude}
\end{equation}
which is minimized over damping ratio $|\zeta_{\overline\lambda}| \le 1$ by 
selecting 
\begin{equation}
\zeta_{\overline\lambda} = 1. 
\label{Eq_extremal_mode_damping_optimal}
\end{equation}
Note that the magnitude of the roots (associated with the extremal eigenvalues) becomes, from Eqs.~\eqref{eq_Lemma_3_proof_mag_beta1_0}, and \eqref{Eq_extremal_mode_general_magnitude},   
\begin{equation}
    |z_{\overline\lambda}| = |z_{\underline\lambda}| = \sqrt{\beta_2}
    = 
    \frac{ \hat{\alpha}(\overline\lambda-\underline\lambda)}{4}
    .
    \label{eq_Lemma_3_proof_mag_beta1_02_}
\end{equation}

\vspace{0.05in}
\noindent 
{\bf{Step 4}} is to find the optimal A-DSR gains $\hat{\alpha}$ and $\beta_2$. 
Substituting $\zeta_{\overline\lambda} = 1$ from Eq.~\eqref{Eq_extremal_mode_damping_optimal} into  Eq.~\eqref{Eq_extremal_mode_damping_exprsn}, results in
\begin{equation}
\begin{aligned}
    \hat{\alpha}\overline\lambda
    & = (1+\beta_2) +2 \sqrt{\beta_2}  \\
     \hat{\alpha}\underline\lambda
    & = (1+\beta_2)-2 \sqrt{\beta_2}.
    \end{aligned}
    \label{Eq_extremal_mode_damping_exprsn_2}
\end{equation}
Dividing the two equations to eliminate $\hat{\alpha}$ yields a quadratic equation for $\sqrt{\beta_2}$, the magnitude  of the roots, 
\begin{equation}
\begin{aligned}
    (\overline\lambda - \underline\lambda)
    \beta_2 
   -2  
   (\overline\lambda + \underline\lambda) \sqrt{\beta_2}
   +(\overline\lambda - \underline\lambda) & =0,
       \end{aligned}
    \label{Eq_extremal_mode_damping_exprsn_4}
\end{equation}
with solutions
\begin{equation}
\begin{aligned}
 \sqrt{\beta_2} & = 
 \frac{(\overline\lambda+\underline\lambda)\pm2\sqrt{\overline\lambda\underline\lambda}}{(\overline\lambda-\underline\lambda)}.
       \end{aligned}
    \label{Eq_extremal_mode_damping_exprsn_5}
\end{equation}

\noindent 
Since $\overline\lambda>\underline\lambda>0$, the smaller root in Eq.~\eqref{Eq_extremal_mode_damping_exprsn_5} is chosen for maximizing structural robustness, resulting in 
\begin{equation}
    \begin{aligned}
    \sqrt{\beta_2} &= \frac{(\overline\lambda + \underline\lambda) -
 2\sqrt{\overline\lambda\underline\lambda}
 }{(\overline\lambda - \underline\lambda)} 
 ~=
 \frac{(\sqrt{\overline\lambda}-\sqrt{\underline\lambda})}{(\sqrt{\overline\lambda}+\sqrt{\underline\lambda})}
    \end{aligned}
    \label{Eq_extremal_mode_damping_exprsn_6}
\end{equation}
and from Eq.~\eqref{eq_Lemma_3_proof_mag_beta1_02_},  
\begin{equation}
    \begin{aligned}
    \hat{\alpha} &= \frac{4}{(\overline\lambda-\underline\lambda)}\sqrt{\beta_2} ~=  \frac{4}{(\sqrt{\overline\lambda}+\sqrt{\underline\lambda})^2}.
    \end{aligned}
    \label{Eq_extremal_mode_damping_exprsn_7}
\end{equation}
\qed

\begin{Rem}[No outdated-feedback in Robust A-DSR] As $\beta_1=0$ for Robust A-DSR (from Lemma~\ref{robust_ADSR_lemma} in Eq.~\eqref{gains_lemma_for_robust_ADSR}), the outdated-feedback term is zero for maximum robustness in A-DSR based approach for networks with real spectrum. Thus, only momentum term is found to be important for improving both robustness and convergence rate of general networks without loops.

\end{Rem}

\subsubsection{Stability with momentum term only}
\vspace{0.1in}
\begin{Lem}
\label{stable_robust_ADSR_lemma1}
{[Stability of Robust A-DSR]}
Let the  A-DSR parameters $\hat{\alpha}$, $\beta_1$ and $\beta_2$ be selected as in Eq.~\eqref{gains_lemma_for_robust_ADSR} from Lemma~\ref{robust_ADSR_lemma} and let the extremal eigenvalues be distinct, i.e.,
$\overline{\lambda}\neq\underline{\lambda}$. Then, the resulting network with the general A-DSR is stable, i.e.,
the 
roots
($ z_{\lambda_{K,m},i}$, $i \in \left\{1,2\right\})$ 
of characteristic Eq.~\eqref{general_char_eq_normal_form}  
(associated with each eigenvalue $\lambda_{K,m}$ of the pinned Laplacian $K$) have magnitude less than one. 

\end{Lem}

{\bf{Proof}~} 
With the optimal parameters in Eq.~\eqref{gains_lemma_for_robust_ADSR}, 
the damping ratio 
$\zeta_{\lambda_{K,m}}$ of
the 
roots
($ z_{\lambda_{K,m},i}$, $i \in \left\{1,2\right\})$ 
of Eq.~\eqref{general_char_eq_normal_form} 
associated with each eigenvalue $\lambda_{K,m}$ of the pinned Laplacian $K$ is given by 
\begin{equation}
    \begin{aligned}
    \zeta_{\lambda_{K,m}} &= \frac{\hat{\alpha}\lambda_{K,m}-1-\beta_2}{2\sqrt{\beta_2}} , 
    \end{aligned} 
    \label{Eq_lemma_middle_eigvals_1}
\end{equation}

\noindent which makes the damping ratio $\zeta_{\lambda_{K,m}}$  linear in the eigenvalue $\lambda_{K,m}$, and varying between  $\zeta_{\underline\lambda}=-1$ to $\zeta_{\overline\lambda}=1$. This  implies that any eigenvalue between the extremal ones is underdamped, i.e.
\begin{equation}
|\zeta_{\lambda_{K,m}}|<1,\;\forall\;\underline\lambda<\lambda_{K,m}<\overline\lambda  
\label{eq_eigenvalue_distribution}
\end{equation}

\noindent As a result, the magnitude of the roots of the characteristic polynomial for $\lambda_{K,m}$ is 

\begin{equation}
    |z_{\lambda_{K,m},1}| = |z_{\lambda_{K,m},2}|= \sqrt{\beta_2} = \frac{\sqrt{\overline\lambda} - \sqrt{\underline\lambda}}{\sqrt{\overline\lambda} + \sqrt{\underline\lambda}} < 1,\; \forall\; \overline \lambda > \underline \lambda > 0, 
    \label{eq_root_location_robust_ADSR}
\end{equation}

\noindent which shows that the roots are strictly within the unit circle resulting in stability.   

\qed

\vspace{0.1in} 
\begin{Rem}[Balanced structural robustness] From Eq.~\eqref{eq_root_location_robust_ADSR}, all the roots 
of the characteristic equation in Eq.~\eqref{general_char_eq_normal_form},  
associated with the Robust A-DSR, have the same magnitude and lie on a circle centered at the origin. Therefore, the roots are equally structurally robust, i.e.,  they are 
equidistant from the unit circle.
Thus, the A-DSR with  optimal  parameters, as in Eq.~\eqref{gains_lemma_for_robust_ADSR} from Lemma~\ref{robust_ADSR_lemma}, 
leads to balanced structural robustness in networks with real spectrum. 
\end{Rem}

\vspace{0.1in}
\section{Results and Discussion}
This section comparatively evaluates  the Optimal no-DSR and the Robust A-DSR approaches using simulation results for an example network's structural robustness and convergence rate during transition. Additionally, the improvements in %the structural robustness and 
convergence rate with the Robust A-DSR are validated with an experimental system.

\vspace{0.1in}

\subsection{Simulation results}
\label{simulation-section}

\vspace{0.1in}

\subsubsection{Example transition problem}
\begin{figure}[!ht]
   \centering
        \includegraphics[width=.80\columnwidth]{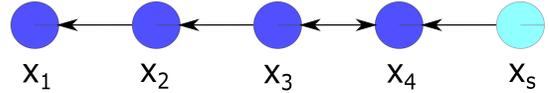} 
    \caption{ Graph of example network  with four agents ($n=4$).  Non-source agents are $X_i, 1 \le i\le 4$, and the source agent is $X_s$. The edge between  agents $X_4$ (the agent with source input) and $X_3$ is undirected, the others are directed.}
    \label{pathgraph} 
\end{figure}

The  network considered here has four  agents ($n=4$) represented by nodes $X_i$, where $1\le i \le 4$,  with node  connectivity represented by the graph  in  Figure~\ref{pathgraph}. Note that the eigenvalues of the given network's Laplacian are real, however the underlying graph is not strongly connected. Moreover, the graph (even without the source $X_s$) is  not balanced.

\noindent 
The virtual source agent $X_s$ determines the desired consensus value for the network and is connected to the agent  $X_4$, i.e. the leader. The connecting edges are all directed in the non-source graph network, except for the undirected edge between the leader $X_4$ and follower agent $X_3$ which makes the graph Laplacian asymmetric.  
The system dynamics with no-DSR for the example network, is given by  Eq.~\eqref{system_non_source},  with the pinned-Laplacian $K$ and $B$ given as

\begin{equation}
    \begin{aligned}
  &K =  \left[
\begin{array}{c c c c}
1  &  -1 & 0 & 0  \\
0 & 1 & -1 & 0 \\ 
0 & 0 & 1 & -1 \\
0 & 0 & -1 & 2 \\
\end{array}
\right] 
&B = \left[
\begin{array}{c}
      0\\
      0\\
      0\\
      1\\
\end{array}
\right].
    \end{aligned}
    \label{pathgraph_Laplacian}
\end{equation}
As discussed in Section~\ref{Sec_Graphs_Laplacian_potential}, the 
 gradient of the asymmetric Laplacian potential $\Phi_{{\cal{G}}}(\hat{X}) =\hat{X}^TL\hat{X}$ in Eq.~\eqref{system_eq_potential} does not lead to standard neighbor-based update in Eq.~\eqref{system_non_source}, where $L$ is the graph Laplacian (from Eq.~\eqref{eq_def_pinned_K}) and $\hat{X}$ is the state vector including source agent.
 
\vspace{0.1in}

\subsubsection{Optimal no-DSR for example network}
\label{consensus-based-simul-subsection}

\begin{figure}[!t]
\begin{center}
    \includegraphics[width=0.90\columnwidth]{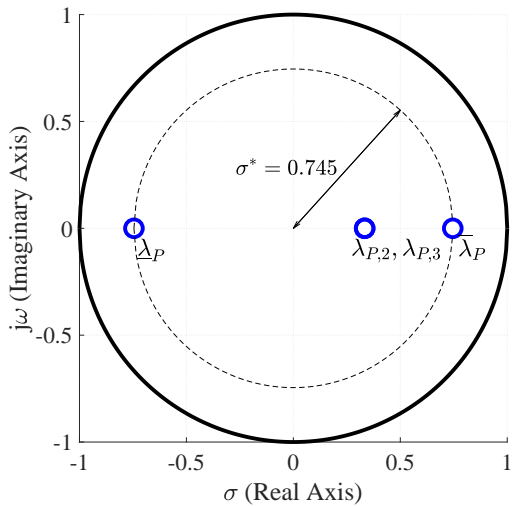}
    \caption{{ Optimal spectral radius ($\sigma^*$) for Optimal no-DSR.} 
     Location of the eigenvalues of matrix P, $\lambda_{P,m}=1-\alpha^* \lambda_{K,m}$  with optimal update gain $\alpha^*$ from Eq.~\eqref{optimal_updategain_pathgraph}. The  spectral radius  with Optimal no-DSR is  $\sigma^*= 0.745$, 
     as in Eq.~\eqref{minimum_spectral_radius} 
     }
    \label{location_of_roots_robust_consensus}
    \end{center}
\end{figure}

The optimal update gain $\alpha^*$ 
from Eq.~\eqref{update_gain_for_max_robust}, 
for minimum spectral radius $\sigma(P)=\sigma(P^*)$, is determined using the extremal eigenvalues $\overline\lambda = 2.618$ and $\underline\lambda = 0.382$ of the pinned-Laplacian $K$ in  Eq.~\eqref{pathgraph_Laplacian}, 
{ using Eq.~\eqref{update_gain_for_max_robust}, } as
\begin{equation}
    \alpha^* = \frac{2}{\overline\lambda + \underline\lambda} = \frac{2}{2.6180 + 0.3820} = 0.6667. 
    \label{optimal_updategain_pathgraph}
\end{equation}
The  measure of structural robustness $d^*$ 
with Optimal no-DSR is, 
{from Eq.~\eqref{eq_optimal_no_dsr_robustness}, }
\begin{equation}
    d^* = 1 - \sigma^* = 0.255, 
\end{equation}
\noindent with the optimal spectral radius $\sigma^* = 0.745$ (from Eq.~\eqref{minimum_spectral_radius}), as illustrated in  Figure~\ref{location_of_roots_robust_consensus}.

\begin{figure}[!ht]
  \begin{center}
    \includegraphics[width=0.75\columnwidth]{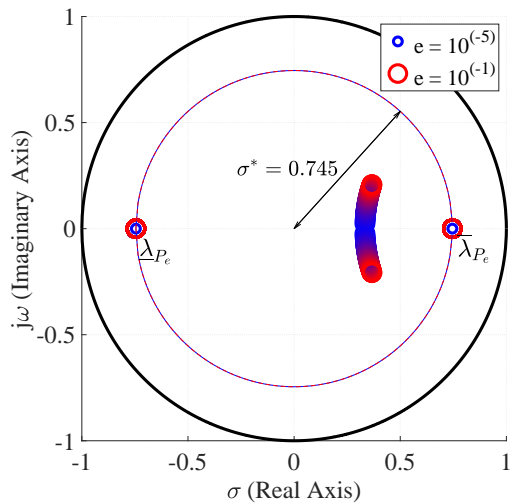}
    \caption{Perturbation $e$ can lead to loss of real spectrum but stability is still structurally robust,  i.e., stability is maintained for small perturbations $e$. Location of the eigenvalues $\lambda_{P_{e}} = 1 - \alpha^* \lambda_{K_{e}}$  of matrix $P_e=\textbf{I}_4-\alpha^* K_e$, where  $K_e$ from Eq.~\eqref{Eq_Laplacian_with_Error} has a perturbation term $e$, which varies from $10^{(-5)}$ (blue) to $10^{(-1)}$ (red). 
    }
    \label{Fig_roots_for_uncertainK}
    \end{center}
\end{figure}

\begin{figure}[!t]
\begin{center}
    \includegraphics[width=0.9\columnwidth]{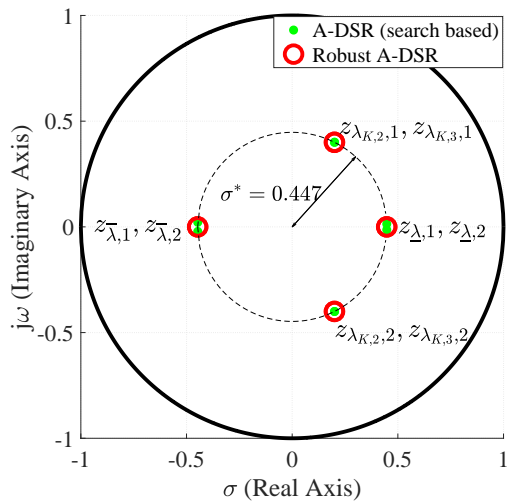}
    \caption{
     Optimal spectral radius ($\sigma^*$) with Robust A-DSR. 
    Location of the roots of characteristic polynomials with Robust A-DSR as in Lemma~\ref{robust_ADSR_lemma} for each eigenvalue $\lambda_{K,m}$ of pinned-Laplacian $K$ in Eq.~\eqref{pathgraph_Laplacian}. The
   spectral radius  with Robust A-DSR is  $\sigma^*=0.447$, an improvement of $40\%$ compared to Optimal no-DSR. 
   The spectral radius with search-based A-DSR is similar to the Robust A-DSR, with similar root locations.
    }
    \label{location_of_roots_critically_damped_ADSR}
    \end{center}
\end{figure}

To assess the transition response, a simulation was  performed with the virtual agent's state $X_s$ changing from an initial value $X_s[k] = x_i$ for all $k<0$ to a final value $X_s[k] = x_f$ for all
 $0\le k$. 
It was assumed that the non-source agents are initially at consensus, i.e., 
$X[0] =  x_i {\textbf{1}}_n$.
With the update gain from Eq.~\eqref{optimal_updategain_pathgraph}, the 
simulated response of the Optimal no-DSR method for a change in virtual agent state $X_s$ from $x_i=0$ to $x_f=100$ is shown in Figure~\ref{ADSR_simul}.
The settling time ($T_s$) of the network's response, defined as the time taken for all the agents' states to achieve and remain within $95\%$ of the desired change $x_{\Delta}=x_f-x_i = 100$ in the consensus state was found to be $14$ sampling time periods ($k=14$) from the simulated response.

\begin{Rem}[Structural robustness with real spectrum]
\label{Structural_robustness_real_spectrum}
The  addition of edges (even with small edge weights) could lead to loss of the real spectrum property. Nevertheless, the stability will still be structurally robust (with or without A-DSR) since the roots of the any general polynomial (and the characteristic equations in particular) are continuous in its coefficients. 
To illustrate, the roots of the pinned-Laplacian with a perturbation term $e$ 
\begin{equation}
K_e = \begin{bmatrix}
1 &-1 &0 &0 \\
e &1-e &-1 &0 \\
0 &0 &1 &-1 \\
0 &0 &-1 &2
\end{bmatrix}
\label{Eq_Laplacian_with_Error}
\end{equation}
are continuous with respect to $e$.
The corresponding location of roots of $P_e = \textbf{I}_4-\alpha^* K_e$ with Optimal no-DSR update gain $\alpha^* = 0.6667$ (obtained from Eq.~\eqref{optimal_updategain_pathgraph}) are shown in Figure~\ref{Fig_roots_for_uncertainK} for increasing perturbation $e$. Although, the resulting spectrum is no longer real, the stability is structurally robust, i.e., stability is maintained for small perturbations $e$. 
\end{Rem}

\vspace{0.1in}
\subsubsection{A-DSR improves structural robustness} 
The A-DSR approach in Eq.~\eqref{accelerated_system_non_source_A_DSR} under  Subsection~\ref{Proposed A-DSR}  is used to improve the example network's structural robustness. The spectral radius of the network is minimized over the range of A-DSR parameters $\hat{\alpha},\beta_1$ and $\beta_2$, 

\begin{equation}
  \sigma^*  =\min_{\hat{\alpha},\beta_1,\beta_2}\left[ \max_{m} \left( \max_{1\le i \le 2}  | z_{\lambda_{K,m},i}| \right) \right], 
   \label{Eq_spectral_radius_minimization}
\end{equation}
\noindent where $z_{\lambda_{K,m},i}$ with $i \in \left\{1,2\right\}$ are the  roots 
of the characteristic Eqs.~\eqref{general_char_eq_normal_form} associated with eigenvalue $\lambda_{K,m}$ of the pinned-Laplacian $K$,  and the search space is constrained by the stability conditions in Eq.~\eqref{general_stability_cond}. The optimum parameters for minimum spectral radius, found through a  numerical search, and the resulting performance are tabulated in Table~\ref{table_simul}.
With these optimal parameter selections, the corresponding roots of the characteristic polynomial with A-DSR, in Eq.~\eqref{accelerated_system_non_source_A_DSR}, for each eigenvalue $\lambda_{K,m}$, are shown in Figure~\ref{location_of_roots_critically_damped_ADSR}.  
The optimal spectral radius is given by 
 $ \sigma^* = 0.447$, which is a reduction of $40\%$ when compared to the Optimal no-DSR case for this example network.  
For the same state transition from 
$x_i= 0$  to $x_f=100$ in the consensus state, the corresponding $5\%$
settling time is $7$ sampling time periods ($k=7$), which is a $50\%$ improvement over the Optimal no-DSR case.  Thus, the A-DSR approach improves both the structural robustness and the convergence rate when compared to the Optimal no-DSR case.

\subsubsection{Robust A-DSR's performance similar to A-DSR}
Instead of a numerical search to optimize the parameters as in the A-DSR case, the 
Robust A-DSR, proposed in Subsection~\ref{subsection_robust_convergence_with_ADSR}, yields closed-form expressions for selection of its parameters as in Eq.~\eqref{gains_lemma_for_robust_ADSR}. With the Robust A-DSR, the corresponding roots of the characteristic polynomials in Eq.~\eqref{general_char_eq_normal_form}, for each eigenvalue $\lambda_{K,m}$, are shown in Figure~\ref{location_of_roots_critically_damped_ADSR}. Note that the  roots corresponding to the extremal eigenvalues $\underline{\lambda}, \overline{\lambda}$ are real valued and critically damped, as in  Lemma~\ref{robust_ADSR_lemma}. Furthermore, the other roots of characteristic equation, for intermediate eigenvalues $\lambda$ satisfying $\underline\lambda<\lambda<\overline\lambda$, lie on a circle with radius equal to magnitude of the critically damped extremal modes as shown in Figure~\ref{location_of_roots_critically_damped_ADSR}, 
which follows from Lemma~\ref{stable_robust_ADSR_lemma1}. 
Overall, the spectral radius $\sigma^*$ of the example network, with Robust A-DSR, is equal to the magnitude of the roots, i.e., 
$\sigma^* = \sqrt{\beta_2} = 0.447$.

\begin{figure}[!ht]
   \hspace{-0.5cm}
        \includegraphics[width=.99\columnwidth]{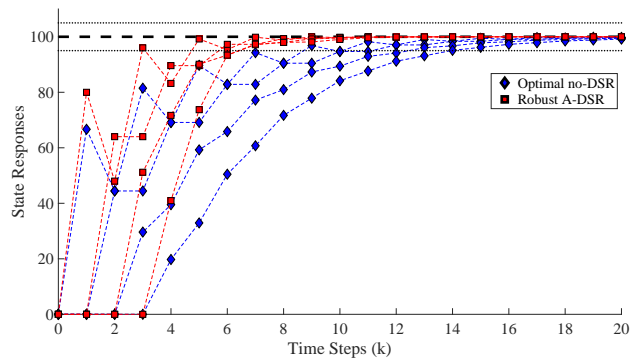}
    \caption{Simulated network state responses with Robust A-DSR (in red) and Optimal no-DSR (in blue, with $\alpha = \alpha^* = 0.6667$), where the Robust A-DSR parameters are chosen as $\hat{\alpha} = 0.80$, $\beta_1 =0$ and $\beta_2=0.20$ from Eq.~\eqref{gains_lemma_for_robust_ADSR} for $\overline\lambda = 2.618$ and $\underline\lambda = 0.382$, showing the settling time $T_s = 7$ sampling time periods ($50\%$ improvement w.r.t. Optimal no-DSR method $T_s = 14$ sampling time periods) 
    }
    \label{ADSR_simul} 
\end{figure}

The performance of the Robust A-DSR is similar to the optimized search-based A-DSR (see Table~\ref{table_simul}).
In particular, the spectral radius of $ \sigma^* = 0.447$ with Robust A-DSR is smaller by $40\%$ when compared to $\sigma(P^*)=0.745$ with the Optimal no-DSR method (see Table~\ref{table_simul}), thus improving the structural robustness.
Additionally, the settling time $T_s$  with Robust A-DSR  was found to be $7$ sampling time periods from the simulation result (which corresponds to a $50\%$ improvement in convergence rate) as shown in Figure~\ref{ADSR_simul}.

{\tiny
\begin{table}[!ht]
\caption{{Simulation results for minimizing (min of) Spectral Radius ($\sigma$) and Settling Time ($T_s$): Comparison of robustness ($\sigma$) \& convergence rates ($T_s$) of network responses using Optimal no-DSR (Eq.~\eqref{system_non_source}), A-DSR (Eq.~\eqref{accelerated_system_non_source_A_DSR}), Nesterov-update (Eq.~\eqref{accelerated_system_non_source}), and the Outdated-feedback and the Momentum  methods}
}
\centering
\begin{tabular}{c c c c c c c}
	\hline\hline % 
		\\    
		Method & min  & $\hat{\alpha}$ &$\beta_1$ &$\beta_2$ & $\sigma$ &$T_s (k)$  \\
		& of   &    \\
		[0.5ex]
		\hline \hline% 
		\\
	\textbf{Robust} & &0.80 &0 &0.20 & 0.4472 &7  \\ 
	 \textbf{A-DSR}& & & & &  &  \\ 
		\hline 
	\textbf{A-DSR} &$ \sigma$ &0.7997 &0.0002 &0.2005 &0.4472 &7  \\
	\textbf{}&$ T_s$ &0.6303 &0.2376 &0.3868 &0.6634 &6 \\
		\hline
			\textbf{Momentum} &$ \sigma$ &0.7995 &0 &0.2006 &0.4479 &7 \\
	&$ T_s$ &0.8388 &0 &0.2347 &0.4845 &6 \\
		\hline
	\textbf{Nesterov} &$ \sigma$ &0.4830 &0.3992 &0.3992 &0.5706 &11 \\
	\textbf{-update}&$ T_s$ &0.5212&0.4684 &0.4684 &0.7599 &7 \\
		\hline
	\textbf{Outdated} &$ \sigma$ &0.9638 &-0.1414 &0 &0.5973 &8 \\
	\textbf{-feedback}&$ T_s$ &1.0874 &-0.1881 &0 &0.7318 &6\\
	\hline \\
	\textbf{Optimal } & &0.6667 &0 &0 &0.745 &14 \\  
		\textbf{no-DSR}& & & & & &\\
		\hline
		\hline  
\end{tabular}
\label{table_simul}
\end{table}
}

\begin{Rem}[Momentum term  $\beta_2$ and 
settling time $T_s$] 
For the Robust A-DSR approach, 
the settling time $T_s$ can be estimated analytically 
in terms of the momentum term $\beta_2$. 
Since all the roots of the characteristic equation in Eq.~\eqref{eq_root_location_robust_ADSR} have the same magnitude,  the dynamics associated with the under-damped roots of the Robust A-DSR converge faster than critically-damped, 
real-valued roots $\sqrt{\beta_2}$.
The corresponding real-valued continuous-time roots $s_{cont}$ are at $s_{cont} = (\ln{\sqrt{\beta_2}})/\delta_t$, which can be used to predict the $5\%$ settling time $T_s$ as (in number of sampling time periods) 
\begin{equation}
    T_{s} \approx \frac{5}{|s_{cont}|\delta_t} ~
    ~ \frac{5}{|\ln{\sqrt{\beta_2}}|} ~= 6.2, 
    \label{Eq_rem_Ts_1}
\end{equation}
which matches the simulation-based value of $7$ sampling time periods. Thus, a larger momentum term $\beta_2$ results in faster settling.
\end{Rem}

 In summary, the Robust A-DSR approach provides similar improvements as with the general A-DSR approach,  in both the structural robustness and the convergence rate when compared to the
Optimal no-DSR approach. The advantage of  the Robust A-DSR approach is that it provides an analytical approach for  selecting the control parameters instead of the numerical search  with the general A-DSR.

\subsubsection{Comparison of constrained accelerated approaches}
Although constrained, the Robust A-DSR (with $\beta_1=0$)  outperforms both the Nesterov-update method (with $\beta_1=\beta_2=\beta$) as well as the Outdated-feedback method (with $\beta_2=0$). 
Optimal parameters for the 
Nesterov-update as well as the Outdated-feedback methods 
were also found using the same optimization in Eq.~\eqref{Eq_spectral_radius_minimization}
with the additional constraints $\beta_1=\beta_2=\beta$ for Nesterov-update method and $\beta_2=0$ for Outdated-feedback method. 
The search space for parameters were constrained as in Corollary~\ref{corrolary_Acc_methods}. 
The optimal parameters of Nesterov-update and Outdated-feedback methods and the performance are provided in Table~\ref{table_simul}. 
When compared to the Optimal no-DSR case, 
the Nesterov-update improves the spectral radius by $23.4\%$ which is less than the improvement of $40\%$ with the Robust A-DSR approach. The Outdated-feedback method also improves the
spectral radius when compared to the no-DSR case, but the improvement ($19.9\%$) is even smaller than the Nesterov-update case with $23.4\%$. Similarly, the settling time improvement of $50\%$ with Robust A-DSR when compared to Optimal no-DSR is larger than the improvement of $21.43\%$ with the Nesterov-update and $42.9\%$ improvement with the Outdated-feedback.
Thus,  while the Robust A-DSR is constrained, it still matches the performance of the general optimal A-DSR, and  outperforms both the Nesterov-update method as well as the Outdated-feedback method.

\begin{Rem}[Outdated-feedback versus momentum] 
When simultaneously improving both the structural robustness and the convergence rate, of the two components of the A-DSR, the momentum component (associated with $\beta_2$) has more significant impact than the outdated-feedback component (associated with $\beta_1$). 
\end{Rem}

\vspace{0.1in}
\subsubsection{Convergence improvement without structural robustness} The above results focused on increasing both the structural robustness and convergence rate. However, the parameters of the accelerated update methods can be chosen purely for optimizing the convergence rate (i.e. minimizing the settling time $T_s$). The resulting optimized parameters (found through a numerical search) and the performance are quantified in Table~\ref{table_simul}.

The accelerated methods achieve smaller settling time $T_s$ when the parameters are optimized for achieving a faster convergence rate. 
For instance, the settling time $T_s$ with A-DSR (search based) improves to $6$ sampling time periods (see Table~\ref{table_simul}), which is faster than Robust A-DSR  and Nesterov-update each taking $7$ sampling time periods, and an improvement of $57.1\%$ over the Optimal no-DSR case. However, this improvement in settling time $T_s$ is accompanied by a decrease in structural robustness of the network. For example, with A-DSR parameters selected for fast convergence, the  spectral radius $\sigma$ increased to $\sigma = 0.6237$ from  $\sigma=\sigma^*=0.4472$ for the case when the parameters were selected to maximize bot the structural robustness and convergence rate. Among the other accelerated approaches, the Momentum method  also achieves the same settling time of $6$ sampling time periods as the A-DSR case, indicating the importance the momentum term in improving the convergence rate of the given example network. A similar loss in structural robustness is seen with the Momentum and Outdated-feedback approaches when the parameters are optimized purely for faster convergence rate, as seen in Table~\ref{table_simul}. The loss in structural robustness (for this example) is more with the  Outdated-feedback than with the Momentum method.

The simulation results show that the network's convergence-rate alone can be improved with the general A-DSR further than that achieved with Robust A-DSR. However, this increase in convergence-rate alone involves a loss in structural robustness. Moreover, the A-DSR parameters are found using a numerical search method. 

In contrast, the parameters of the Robust A-DSR can be found analytically and it achieves similar convergence rate as the A-DSR optimized for convergence-rate alone. Moreover, the performance improvement with the Robust A-DSR (as well as the A-DSR), in terms of both the 
structural robustness 
and the rapidity of transition, is better than the  performance with the standard no-DSR consensus method.

\subsection{Experimental results}
\label{experimental-section}

A mobile-bot network is used for experimental evaluation of the proposed A-DSR approach. 

\subsubsection{System description}
\begin{figure}[!th] 
\begin{center}
\includegraphics[width=0.95\columnwidth]{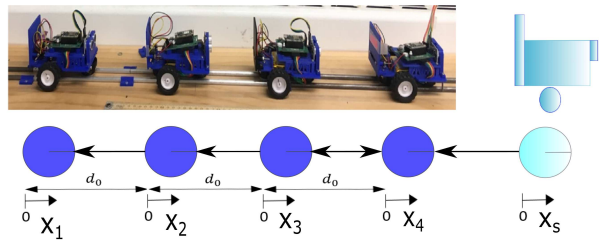}
    \caption{
 Experimental test bed consisting of four mobile-bot agents moving in a straight line, with the same  connectivity  as  in the example simulation network in Figure~\ref{pathgraph}. Each $i^{th}$ agent's state is its displacement $X_i$ from its initial position.
     }
    \label{bot_network}
\end{center}
\end{figure}

The experimental setup consists of four mobile-bot agents that move in a straight line. The network connectivity is the same as in the simulation example. The bots aim to maintain a spacing of $d_o$ between them, and the state $X_i$ of each bot $i$ is defined as the displacement from the initial equally-spaced configuration, as shown in Figure~\ref{bot_network}. The virtual source input $X_s$ determines the desired position of the network.

\subsubsection{Bot's update computation}
The desired displacement $X_i[k+1]$ at the next time step $k+1$ is computed using local relative-distance measurements available at time step $k$ by each bot $i$ using 
distance sensors (Ultrasonic HC-SR04 to the front, and Infrared GP2Y0A21YK at the back).
These  measurements of each bot $i$ include  
\begin{equation}
     X_{f,i}[k]=(X_{i+1}[k]-X_i[k]) + d_0, 
\end{equation} 
the relative displacement w.r.t. the front bot  $i+1$ (which is $X_s$ for leader bot $i=4$), and 
\begin{equation}
X_{b,i}[k] = d_0 - (X_{i-1}[k]-X_i[k]), 
\end{equation}
 the relative displacement 
w.r.t. the back bot $i-1$ where $2< i< 4$,  and $d_0$ is the desired offset distance between the bots in the experimental setup. These relative-distance measurements 
 ($ X_{f,i}[k], X_{b,i}[k] $) are used to determine the neighbor information needed to evaluate the update law, i.e., to obtain $K_iX[k]$, where $K_i$ is the $i^{th}$ row of the pinned-Laplacian in Eq.~\eqref{pathgraph_Laplacian}. For example, 
 \begin{equation} 
 \begin{aligned}
 K_i X[k] & =   K_{i,i+1} (X_{i}[k]-X_{i+1}[k])  \\
   & \qquad + K_{i,i-1}  (X_{i}[k] - X_{i-1}[k])  \\
 & =   K_{i,i+1} (d_0 - X_{f,i}[k]) + K_{i,i-1}  (X_{b,i}[k] - d_0). 
  \end{aligned}
 \label{eq_relative_positon_measurements}
 \end{equation}
Thus, the relative-distance  measurements ($ X_{f,i}[k], X_{b,i}[k] $) at time step $k$ 
enable each bot $i$ to compute its update, 
i.e., to find the desired position $X_i[k+1]$ at the next time step  according to Eq.~\eqref{accelerated_system_non_source_single}, where parameters $\beta_1$ and $\beta_2$ are zero for the no-DSR case.

\subsubsection{Bot's feedback control}
\begin{figure}[!ht]
\centering
      \includegraphics[width=\columnwidth]{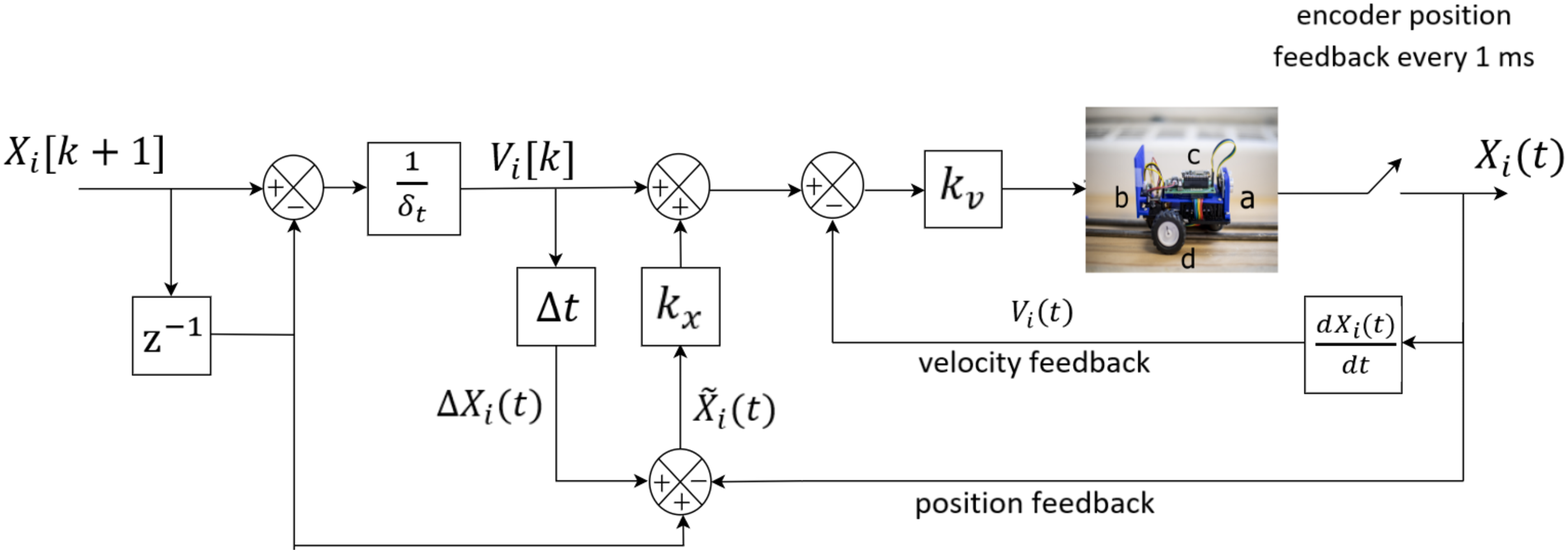} 
\caption{
{Each $i^{th}$ bot's control system includes: 
a) distance sensors to the front and, b) back , c) micro-controller for on-board computation, and d) wheels with magnetic encoders on motors to estimate each bot's displacement, $X_{i}(t)$. 
To ensure that the bot achieves $X_i[k+1]$, an inner-loop controller with gain $k_v$ to track desired velocity $V_i[k]$ in Eq.~\eqref{Eq_desired_Velocity_at_tk} and an outer-loop controller with gain $k_x$ for position error ($\Tilde{X}_i(t)$ in Eq.~\eqref{Eq_position_error}) are implemented. 
}
}
\label{SensorsOnBot} 
\end{figure}

Each $i^{th}$ bot's controller aims to match its 
state (displacement)
$X_{i}(t)$ to be the desired  state $X_i[k+1]$  by the next time step, i.e., when time $t=t_{k+1}$.
This is accomplished using a velocity-feedback inner-loop and a 
position-feedback outer-loop,  as shown in  Figure~\ref{SensorsOnBot}, using 
measurements of the agent state $X_i(t)$ from magnetic encoders on each bot $i$.

\noindent 
In particular, the desired velocity for the time period $[t_k, t_{k+1})$ is computed as 
\begin{equation}
V_i[k] =  \frac{X_i[k+1]-X_i[k]}{\delta_t}, 
\label{Eq_desired_Velocity_at_tk}
\end{equation} 
where $\delta_t$ is the discrete time step (in seconds) for the update method. 
The desired velocity $V_i[k]$ is then tracked using an inner-loop controller with gain $k_v$ as shown in Figure~\ref{SensorsOnBot}. Additionally, an outer-loop feedback with gain  $k_x$ is used to correct for position error ($\Tilde{X}_i(t)$) at any time $t\in [t_k, t_{k+1})$, determined as

\begin{equation}
    \Tilde{X}_i(t) = (X_i[k]+\Delta X_i(t)) - X_i(t), 
    \label{Eq_position_error}
\end{equation}

\noindent where $\Delta X_{i}(t) = V_i[k] (t-t_k) = V_i[k]\Delta t$, as  shown in Figure~\ref{SensorsOnBot}.

The selection of position transition magnitude for the experiment was based on velocity limits of $20$ cm/s for the bots. The initial position was $x_i=0$, and the final position was $x_f =100$ cm. Therefore the sampling time period $\delta_t$ was chosen as $4$ s to ensure that the bots could meet the maximum position transitions of $80$ cm in one sampling-time period $\delta_t$, seen in simulations in Figure~\ref{ADSR_simul}, with the bot's feedback gains $k_v =5$ and $k_x = 1$.

\begin{table}[!t]
	\caption{{ Experimental results. Comparison of convergence rate in position responses with Robust A-DSR and Optimal no-DSR for multi-agent network in Figure~\ref{bot_network}, using settling time $T_s$.}}
	\centering
\begin{tabular}{c c c}
	\hline\hline % 
		\\ 
		Method & Trial  &$T_s (k)$  \\%  
[0.5ex]
		\hline \hline% 
		\\
	\textbf{Robust A-DSR} & Trial 1 &11  \\ 
	       &Trial 2  &11  \\ 
	 ($\hat{\alpha}=0.80,\beta_1=0, \beta_2=0.20$)& Trial 3  &10  \\ 
	 &Trial 4  &  11\\ 
	 &Trial 5  &  11\\ 
	 &Trial 6  &  10\\ 
	 &Trial 7  &  9\\ 
		\hline 
		\bf{Mean Response} & &\bf{10} \\
		\hline  \\
	\textbf{Optimal no-DSR} &Trial1  &17 \\  
	&Trial2 & 16\\
	($\alpha = \alpha^*=0.67$)&Trial3 & 16\\
	&Trial4 & 15\\
	&Trial5 & 15\\
	&Trial6 & 18\\
	&Trial7 & 15\\
		\hline%  
		\bf{Mean Response} & &\bf{16} \\
		\hline %  
		\hline
\end{tabular}
	\label{table:ExpDSR} %  
\end{table}

 \begin{figure}[!t]
	\hspace{-0.5cm}
	\includegraphics[width=0.99\columnwidth]{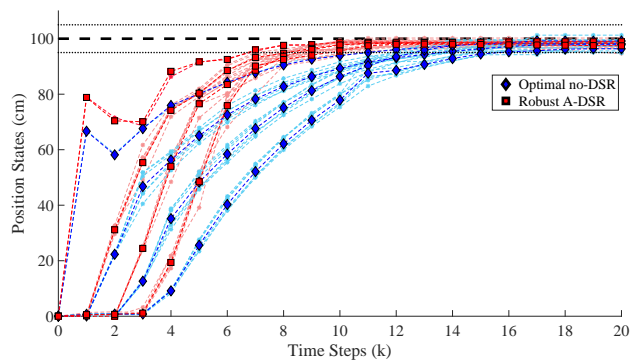}
	\caption{Experimental position responses over 7 trials (in lighter shade) and their mean (in dark lines) in the experiments comparing the Optimal no-DSR (in blue) and Robust A-DSR (in red) methods for fast convergence. The experiments on average show an improvement with Robust A-DSR of $37.5\%$ in $T_s$ (from 16 time steps to 10 time steps).
	}
	\label{averageExp_withinput} 
\end{figure}

\subsubsection{Convergence rate improvement}
The improvement in convergence rate of transition response in the example network with Robust A-DSR, over  Optimal no-DSR, is evaluated through the experimental mobile-bot network.

A transition in desired position (defined using virtual source $X_s$) from $x_i=0$ cm to $x_f=100$ cm, similar to simulations, is implemented on the mobile-bot network. Each bot, initially in consensus with position zero, responds as the transition information propagates through the bot network (in Figure~\ref{bot_network}). This state transition is implemented using Optimal no-DSR and Robust A-DSR, with parameters given in Table~\ref{table_simul}, and the observations of convergence rates from seven trials (with both the approaches) are tabulated in Table~\ref{table:ExpDSR}. The position responses of the bots during the transition are plotted in Figure~\ref{averageExp_withinput}, for each of the seven trials with Optimal no-DSR (in light blue) and Robust A-DSR (in light red). The mean responses for both approaches, obtained from averaging over the seven trials, are also shown in Figure~\ref{averageExp_withinput}.

\indent 
Robust A-DSR shows  improvement  in convergence rate of the bot network's transition response, improving the settling time (within $5\%$ of the final position) by $4$ to $9$ time periods ($27$\% to $50$\%), when compared with Optimal no-DSR, similar to that observed in simulations. The mean response converges $6$ time periods faster with Robust A-DSR (an improvement of $37.5\%$) when compared with Optimal no-DSR, see Table~\ref{table:ExpDSR}.
Thus, the convergence rate improvements observed in simulations with Robust A-DSR, with analytically determined parameters, over Optimal no-DSR are verified with similar results from experimental studies of position transition in the mobile-bot network.

\section{CONCLUSIONS}
\label{conclusion-section}
The article introduced an accelerated delayed self reinforcement (A-DSR) approach, based on local potential, for improving the structural robustness and convergence rate beyond the limits of standard consensus-based networks. Of the two terms in the accelerated approach, it was shown that the momentum term has substantially more impact when compared to the outdated-feedback term for improving convergence rate and robustness in networks with real spectrum. A Robust A-DSR approach was developed, with analytical expressions for its parameters, that closely matches the performance of the general A-DSR approach, which alleviates the  need for numerical search when selecting parameters of the general A-DSR. Moreover, experimental results verified the improved convergence rate with Robust A-DSR over Optimal no-DSR. 

The A-DSR approach, presented in this work, assumes scalar gains 
for the outdated-feedback and momentum terms, which can be extended in future work by using different, possibly nonlinear or time-varying gains for each agent in the network. Further, the proposed Robust A-DSR approach, can be used to accelerate convergence and improve performance of networks with uncertainty, for instance, distributed sensing in presence of communication delays, operation of multi-agent networks with a human-in-the-loop where the human or network model is uncertain, and transporting flexible structures with uncertain stiffness values using mobile bots. Further work is needed to explore the suitability of the Robust A-DSR for these applications.

\bibliographystyle{unsrt}
\bibliography{ASMEold}

\end{document}